\documentclass[3p]{elsarticle}
\usepackage[T1]{fontenc}
\usepackage[latin9]{inputenc}
\usepackage{color}
\usepackage{array}
\usepackage{bm}
\usepackage{multirow}
\usepackage{amsmath}
\usepackage{amssymb}
\usepackage{graphicx}
\usepackage{wasysym}
\usepackage[unicode=true,
 bookmarks=true,bookmarksnumbered=false,bookmarksopen=false,
 breaklinks=false,pdfborder={0 0 0},pdfborderstyle={},backref=false,colorlinks=true]
 {hyperref}

\makeatletter

\providecommand{\tabularnewline}{\\}
\newcommand{\lyxdot}{.}

\biboptions{sort&compress}

\usepackage{xcolor}
\usepackage{textcomp}
\usepackage{hyphenat}
\usepackage{makecell} 
\usepackage[]{lmodern} 
\usepackage{pifont}
\usepackage{listings}
\makeatother

\usepackage{listings}

\begin{document}

\title{\textsf{MSGCorep}: A package for corepresentations of magnetic space
groups}

\author[bit1,bit2]{Gui-Bin Liu\corref{cor}}

\ead{gbliu@bit.edu.cn}

\cortext[cor]{Corresponding author}

\author[buct]{Zeying Zhang}

\author[bit1,bit2]{Zhi-Ming Yu}

\author[bit1,bit2]{Yugui Yao\corref{cor}}

\ead{ygyao@bit.edu.cn}

\address[bit1]{Centre for Quantum Physics, Key Laboratory of Advanced Optoelectronic
Quantum Architecture and Measurement (MOE),\\
School of Physics, Beijing Institute of Technology, Beijing 100081,
China}

\address[bit2]{Beijing Key Laboratory of Nanophotonics and Ultrafine Optoelectronic
Systems, School of Physics,\\
Beijing Institute of Technology, Beijing 100081, China}

\address[buct]{College of Mathematics and Physics, Beijing University of Chemical
Technology, Beijing 100029, China}
\begin{abstract}
Motivated by easy access to complete corepresentation (corep) data
of all the 1651 magnetic space groups (MSGs) in three-dimensional
space, we have developed a Mathematica package \textsf{MSGCorep} to
provide an offline database of coreps and various functions to manipulate
them, based on our previous package \textsf{\textsf{\mbox{}SpaceGroupIrep}}.
One can use the package \textsf{MSGCorep} to obtain the elements of
any MSG and magnetic little group, to calculate the multiplication
of group elements, to obtain the small coreps at any k-point and full
coreps of any magnetic k-star for any MSG and show them in a user-friendly
table form, to calculate and show the decomposition of direct products
of full coreps between any two specified magnetic k-stars, and to
determine the small coreps of energy bands. Both single-valued and
double-valued coreps are supported. In addition, the 122 magnetic
point groups (MPGs) and their coreps are also supported by this package.
To the best of our knowledge, \textsf{MSGCorep} is the first package
that is able to calculate the direct product of full coreps for any
MSG and able to determine small coreps of energy bands for general
purpose. In a word, the \textsf{MSGCorep} package is an offline database
and tool set for MSGs, MPGs, and their coreps, and it is very useful
to study the symmetries in magnetic and nonmagnetic materials.
\end{abstract}
\begin{keyword}
corepresentation, magnetic space group, magnetic little group, direct
product, Mathematica
\end{keyword}
\maketitle

\global\long\def\vr{\bm{r}}%
\global\long\def\vR{\bm{R}}%
\global\long\def\vk{\bm{k}}%
\global\long\def\vK{\bm{K}}%

\global\long\def\bktwo#1#2{\langle#1|#2\rangle}%

\global\long\def\bkthree#1#2#3{\langle#1|#2|#3\rangle}%

\global\long\def\ket#1{|#1\rangle}%
\global\long\def\bra#1{\langle#1|}%

\global\long\def\ave#1{\langle#1\rangle}%

\global\long\def\veps{\varepsilon}%

\global\long\def\tr{\mathcal{T}}%

\global\long\def\cosets{/\!\!/}%

\global\long\def\herring#1{\vphantom{#1}^{H}\!#1}%

\section*{Program summary}

\noindent \textit{Program title}: \textsf{MSGCorep}

\noindent \textit{Developer's respository link}: \url{https://github.com/goodluck1982/MSGCorep}

\noindent\textit{Licensing provisions}: GNU General Public Licence
3.0

\noindent\textit{Distribution format}: tar.gz

\noindent\textit{Programming language}: Wolfram

\noindent\textit{External routines/libraries used}: \textsf{SpaceGroupIrep}
(\url{https://github.com/goodluck1982/SpaceGroupIrep})

\noindent\textit{Nature of problem}: The package \textsf{MSGCorep}
provides offline database and tools for easy access to 1651 magnetic
space groups, 122 magnetic point groups, and their corepresentations.
\textsf{MSGCorep} is the first package that is able to calculate the
direct product of full corepresentations for any magnetic space group
and able to determine small corepresentations of energy bands for
general purpose.

\noindent\textit{Solution method}: Corepresentations of a magnetic
group are constructed from the representations of the maximal unitary
subgroup of the magnetic group. Based on the complete representation
data of 230 space groups provided by \textsf{SpaceGroupIrep} package,
complete corepresentation data of 1651 magnetic space groups are derived
in \textsf{MSGCorep} package.

\section{Introduction}

It's well known that group theory and symmetries play important roles
in physics, especially in the booming topological physics in the recent
decade \citep{Wieder_Bernevig_2022_7_196__Topological,Wieder_Bernevig_2018_361_246_1705.01617v1_Wallpaper,Jadaun_Banerjee_2013_88_85110__Topological,Elcoro_Bernevig_2020_102_35110__Application,Yu_Yao_2022_67_375_2102.01517_Encyclopedia,Wu_Wan_2021_104_45107_2102.09515_Symmetry}.
An typical example is the application of space groups (SGs) and their
representation theory in symmetry indicator method or topological
quantum chemistry method to classify and search symmetry-protected
band topology in nonmagnetic materials \citep{Po_Watanabe_2017_8_50__Symmetry,Kruthoff_Slager_2017_7_41069__Topological,Song_Fang_2018_9_3530_1711.11049v3_Quantitative,Zhang_Fang_2019_566_475__Catalogue,Tang_Wan_2019_15_470__Efficient,Tang_Wan_2019_566_486__Comprehensive,Tang_Wan_2019_5_8725__Topological,Bradlyn_Bernevig_2017_547_298__Topological,Cano_Bernevig_2018_120_266401__Topology,Cano_Bernevig_2018_97_35139__Building,Vergniory_Wang_2019_566_480__complete,Cano_Bradlyn_2021_12_225__Band},
and these two methods have been extended to magnetic space groups
(MSGs) which are needed to describe the symmetries in magnetic materials
\citep{Watanabe_Vishwanath_2018_4_8685__Structure,Xu_Bernevig_2020_586_702__High,Elcoro_Bernevig_2021_12_5965_2010.00598_Magnetic}.
Consequently, more and more interests are focused on MSGs and their
applications in magnetic topological materials \citep{Watanabe_Watanabe_2018_97_165117__Lieb,Liu_Yao_2022_105_85117__Systematic,Zhang_Yao_2022_105_104426_2112.10479_Encyclopedia,Tang_Wan_2021_104_85137_2103.08477_Exhaustive,Tang_Wan_2022___2201.09836_Complete,Ono_Shiozaki_2021_3_23086__Z2,Bernevig_Beidenkopf_2022_603_41__Progress}.

The term ``magnetic (space) group'' is also known as ``Shubnikov
(space) group'' originating from A. V. Shubnikov who introduced the
concept of anti-symmetry operation \citep{BCbook}. The anti-symmetry
operation commutes with all spacial operations and can be defined
as the change of an extra two-valued property. The change of color
(black or white) is usually such an anti-symmetry operation in most
general discussions about Shubnikov groups. While in the certain case
of describing magnetic structures, time reversal operation $\tr$
is the anti-symmetry operation which can reverse the direction (or
sign) of magnetic moments. Accordingly, it can be said that MSG is
a certain realization of Shubnikov group. In the case of MSG, $\tr$
is an anti-unitary operation, but it should be emphasized that the
anti-symmetry operation is not necessarily anti-unitary for general
Shubnikov groups \citep{BCbook} and the color-changing operation
is not always the same with $\tr$ \citep{Burns_Glazer_2013____Space}.
Nevertheless, MSGs are often described in terms of changing color
due to historical reasons.

There are in total 1651 MSGs in three-dimensional space which can
be divided into four types. The ordinary 230 SGs (also called Fedorov
groups), which are all unitary, constitute the type-I MSGs, i.e. $M_{{\rm I}}=G$
where $M_{{\rm I}}$ and $G$ stand for type-I MSG and ordinary SG
respectively. It is noteworthy that type-I MSGs can describe the magnetic
structures of ferromagnetic (including ferrimagnetic, the same below)
and antiferromagnetic, but not nonmagnetic (or paramagnetic) materials
\citep{typeI}. Type-II MSGs (230 in total, also called gray SGs)
are of the form $M_{{\rm II}}=G+\tr G$ and can only describe nonmagnetic
materials. Type-III MSGs (674 in total) are of the form $M_{{\rm III}}=H+\tr(G-H)$
in which the equi-translational subgroup $H$ consists of half the
elements in $G$ and is also the maximal unitary subgroup of $M_{{\rm III}}$.
Similar to type-I MSGs, type-III MSGs can describe both ferromagnetic
and antiferromagnetic materials. Type-IV MSGs (517 in total) are of
the form $M_{{\rm IV}}=G+\tr\{E|\bm{t}_{0}\}G$ in which $\{E|\bm{t}_{0}\}$
is a pure translation in Seitz notation with $\bm{t}_{0}$ half a
lattice vector. Type-IV MSGs can only describe antiferromagnetic materials.
Both type-III and type-IV MSGs are also called black-white SGs but
only type-IV MSGs have black-white Bravais lattice due to the existence
of $\tr\{E|\bm{t}_{0}\}$. Alternatively, type-III and type-IV are
called $M_{T}$ and $M_{R}$ respectively by Litvin \citep{Litvin2013MGT,Litvin2016inITA}
and called BW1 and BW2 respectively by Grimmer \citep{Grimmer_Grimmer_2009_65_145__Comments}.

The tabulation of the irreducible representations (ireps) for the
230 SGs have been given in serveral monographs \citep{Kovalev1965,Miller_Love_1967____Tables,ZakCGG,BCbook,CDMLbook},
and the electronic data of them are also available in the \textsf{ISOTROPY}
software suit \citep{isotropy,Stokes_Cordes_2013_69_388__Tabulation},
on the Bilbao Crystallographic Server (BCS) \citep{BCSweb,Aroyo_Wondratschek_2006_A62_115__Bilbao,Elcoro_Aroyo_2017_50_1457__Double},
or in the \textsf{SpaceGroupIrep} package \citep{Liu_Yao_2021_265_107993__SpaceGroupIrep}.
Introducing anti-unitary operations to MSGs makes it necessary to
apply corepresentations (coreps), instead of ireps, to analyze the
symmetries in most physics problems. However, to the best of our knowledge,
the tabulation of irreducible coreps (irreducible is omitted for simplicity
hereinafter) for all MSGs exists only in one book by Miller and Love
\citep{Miller_Love_1967____Tables}, and the tabulation there is not
complete and also difficult to read, as pointed out in Ref. \citep{Elcoro_Bernevig_2021_12_5965_2010.00598_Magnetic}.
Required by the study of magnetic topological quantum chemistry ,
it was not until recent two years that the first complete tabulation
of the coreps for all MSGs were calculated and made publicly available
on BCS \citep{Elcoro_Bernevig_2021_12_5965_2010.00598_Magnetic,Xu_Bernevig_2020_586_702__High}.
BCS is an excellent website for showing crystallographic data. It's
convenient to lookup small coreps of any specified k-point (i.e. wave
vector) for any specified MSG on BCS. However, it's not convenient
to process all coreps of all MSGs ergodically because BCS does not
provide offline tool to do this. Motivated by solving this problem,
we have developed the package \textsf{MSGCorep} to provide offline
database and functions for all coreps of MSGs, based on our homemade
\textsf{SpaceGroupIrep} package for SGs as well as their ireps \citep{Liu_Yao_2021_265_107993__SpaceGroupIrep}.

Generally speaking, the first step of applying the theory of MSG coreps
to analyze energy bands is to determine the coreps of the bands, or
more specifically, the irreducible coreps of the magnetic little group
(MLG), also called ``small coreps'' here, at any required wave vector
$\vk$ of the bands. Similarly, for ordinary SGs this step is to determine
the small representations (reps) of the little groups (LGs) of the
bands. There have been several tools capable of doing this, e.g. \textsf{irvsp}
\citep{Gao_Wang_2021_261_107760_2002.04032_Irvsp}, \textsf{qeirreps}
\citep{Matsugatani_Watanabe_2021_264_107948__qeirreps}, \textsf{IrRep}
\citep{Iraola_Tsirkin_2022_272_108226_2009.01764_IrRep}, and \textsf{SpaceGroupIrep}
\citep{Liu_Yao_2021_265_107993__SpaceGroupIrep}. However, there is
still missing such a tool which can determine the MLG small coreps
for bands. Nevertheless, there is a precursor tool \textsf{MagVasp2trace}
\citep{MagVasp2trace,Xu_Bernevig_2020_586_702__High} which is used
to calculate the characters (i.e. the trace data) of the unitary operations
of the MLGs for bands. In fact, \textsf{MagVasp2trace} was developed
to assist the study of magnetic topological quantum chemistry, i.e.
to prepare the trace data for the ``Check Topological Magnetic Mat.''
(CTMM) function on BCS \citep{Xu_Bernevig_2020_586_702__High,Gao_Wang_2022___2204.10556_Magnetic}.
Although the CTMM has to determine first the small coreps based on
the input trace data, it is designed to analyze the topological properties
using the determined coreps, not to give the small coreps directly.
As a result, the CTMM is not a general tool to determine small coreps
at any k-points. To meet this requirement, we developed the \lstinline!getBandCorep!
function in our \textsf{MSGCorep} package which can determine small
coreps at any k-points based on the trace data from \textsf{MagVasp2trace}.
This is the first tool to do this as we know.

\section{Theory}

\subsection{Basis for corep}

It's well known that a group rep is defined by a homomorphism mapping
to a matrix group. However, the corep of a nonunitary group, say a
magnetic group $M$, is not defined by homomorphism any more. The
corep matrices $D(M)$ obey the following algebra
\begin{equation}
\forall m_{1},m_{2}\in M\ \ \begin{cases}
D(m_{1})D(m_{2})=D(m_{1}m_{2}), & m_{1}\text{ is unitary}\\
D(m_{1})D(m_{2})^{*}=D(m_{1}m_{2}), & m_{1}\text{ is anti-unitary}
\end{cases}\label{eq:corepAlg}
\end{equation}
and the set of matrices $D(M)$ is not necessarily a group. The corep
of $M$ can be induced from an irep (a corep) $\Gamma$ of its unitary
(nonunitary) subgroup $S$. However, this induction procedure is different
from that of ordinary induced rep. To reflect this difference, here
we use the symbol $\Uparrow$ instead of the ordinary $\uparrow$
to describe the induced corep. Supposing that the (left) coset\footnote{All cosets refer to left cosets unless otherwise stated. }
representatives of $S$ are $c_{1},c_{2},\cdots,c_{r}$ with $r=|M|/|S|$
being the index of the subgroup $S$ and $|M|$ being the size of
the set $M$, then for $\forall m\in M$ the induced corep $\Gamma\Uparrow M$
of $M$ is defined as 
\begin{equation}
(\Gamma\Uparrow M)_{\mu\nu}(m)=\begin{cases}
\Gamma(c_{\mu}^{-1}mc_{\nu}), & c_{\mu}^{-1}mc_{\nu}\in S\text{ and }c_{\mu}\text{ is unitary}\\
\Gamma(c_{\mu}^{-1}mc_{\nu})^{*}, & c_{\mu}^{-1}mc_{\nu}\in S\text{ and }c_{\mu}\text{ is anti-unitary}\\
0, & \text{otherwise}
\end{cases}.\label{eq:indcorep}
\end{equation}

In fact, any nonunitary group $M$ can be written as $M=H+AH,$ in
which $H$ is the maximal unitary subgroup of $M$, $A$ is an anti-unitary
coset representative, and the coset $AH$ contains all anti-unitary
elements in $M$. In terms of Eq. (\ref{eq:indcorep}), we have $S=H,$
$\Gamma=\Delta,$ $r=2,$ $c_{1}=E$ (identity element), $c_{2}=A$,
and the obtained induced corep is 
\begin{equation}
(\Delta\Uparrow M)(m)=\begin{bmatrix}\Delta(m) & 0\\
0 & \Delta(A^{-1}mA)^{*}
\end{bmatrix},\ \ (\Delta\Uparrow M)(m')=\begin{bmatrix}0 & \Delta(m'A)\\
\Delta(A^{-1}m')^{*} & 0
\end{bmatrix}
\end{equation}
for $\forall m\in H$ and $\forall m'\in AH$. However, although $\Delta$
is irreducible for $H$, the induced corep $\Delta\Uparrow M$ is
not necessarily irreducible in general. Corep theory shows that only
one irreducible corep (up to equivalence) remains after the reduction
of $\Delta\Uparrow M$ and there are only three cases (a), (b), and
(c) \citep{BCbook}. In both cases (a) and (b), $\Delta(m)$ is equivalent
to $\Delta(A^{-1}mA)^{*}$, i.e. there existing a matrix $N$ satisfying
$\Delta(m)=N\Delta(A^{-1}mA)^{*}N^{-1}$ for $\forall m\in H$, while
in case (c) $\Delta(m)$ is not equivalent to $\Delta(A^{-1}mA)^{*}.$
Denoting the irreducible corep derived from $\Delta$ as $\mathcal{D}\Delta$,
then for $\forall m\in H$ and $\forall m'\in AH$ the results are
\begin{eqnarray}
\text{(a)} & \negthickspace & \mathcal{D}\Delta(m)=\Delta(m),\ \ \mathcal{D}\Delta(m')=\Delta'(m'),\ \ NN^{*}=\Delta(A^{2})\label{eq:corepa}\\
\text{(b)} & \negthickspace & \mathcal{D}\Delta(m)=\begin{bmatrix}\Delta(m) & 0\\
0 & \Delta(m)
\end{bmatrix},\ \ \mathcal{D}\Delta(m')=\begin{bmatrix}0 & -\Delta'(m')\\
\Delta'(m') & 0
\end{bmatrix},\ \ NN^{*}=-\Delta(A^{2})\label{eq:corepb}\\
\text{(c)} & \negthickspace & \mathcal{D}\Delta(m)=(\Delta\Uparrow M)(m)\label{eq:corepc}
\end{eqnarray}
where $\Delta'(m')\equiv\Delta(m'A^{-1})N$. For the matrix $N$ refer
to \ref{sec:matN} for details. The case of a corep $\mathcal{D}\Delta$
can be determined simply by the following equation \citep{BCbook}
\begin{equation}
\sum_{m'\in AH}\chi(m'^{2})=\begin{cases}
+|H| & \text{: case (a)}\\
-|H| & \text{: case (b)}\\
0 & \text{: case (c)}
\end{cases}
\end{equation}
in which $\chi$ is the character of $\Delta$. This is related to
the additional degeneracy of a system when its symmetry group changes
from $H$ to $M.$ In case (a), no additional degeneracy exists, i.e.
the degeneracy keeps unchanged from $H$ to $M$. While in both cases
(b) and (c) additional degeneracy appears, i.e. the degeneracy doubles
from $H$ to $M$. The difference is that in case (b) two copies of
the same irep of $H$ form a corep of $M$ but in case (c) two different
ireps of $H$ form a corep of $M$.

\subsection{Coreps for the 1651 MSGs}

All the 1651 MSGs consist of both unitary groups (type-I) and nonunitary
groups (type-II, -III, and -IV). Although corep is defined for nonunitary
groups, it is still meaningful for unitary groups. Because Eq. (\ref{eq:corepAlg})
becomes $D(m_{1})D(m_{2})=D(m_{1}m_{2})$ for $\forall m_{1},m_{2}\in M$
when no anti-unitary element exists in $M$, in which case the corep
is just an ordinary rep. Accordingly, for the sake of concise and
unified description, we use the term ``corep'' for all MSGs and
MLGs no matter they are unitary groups or not.

Suppose $M$ is an MSG and $M_{\vk}$ is one of its MLGs. $M_{\vk}$
is defined at a certain wave vector $\vk$ and consists of all the
elements in $M$ whose point parts transform $\vk$ to its equivalent
wave vectors, i.e. $M_{\vk}=\{\,Q\,|\,Q\in M\,\&\mathcal{\,P}(Q)\vk\doteq\vk\}$.
This is similar to the LG concept for SG. The point part of the element
$Q$ in $M$ is defined as $\mathcal{P}(Q)=R$ if $Q=\{R|\bm{v}\}$
and $\mathcal{P}(Q)=\mathcal{T}R$ if $Q=\mathcal{T}\{R|\bm{v}\}=\{\mathcal{T}R|\bm{v}\}$,
in which $R$ is a rotation and $\bm{v}$ is a translation vector
(not necessarily a lattice vector). The symbol $\doteq$ means the
equivalence between two k-points, i.e. differing by a reciprocal lattice
vector. And we use the convention $\mathcal{T}\vk=-\vk$ here. $M_{\vk}$
is a subgroup of $M$, and hence $M$ can be written as $M=c_{1}M_{\vk}+c_{2}M_{\vk}+\cdots+c_{r}M_{\vk}$
in which the first coset representative $c_{1}$ is usually set to
$\{E|\bm{0}\}$. Then, the magnetic (wave vector) star of $\vk$ can
be defined as $^{*}\vk=\{\vk_{\mu}\,|\,\mu=1,2,\cdots,r\}$ with $\vk_{\mu}\equiv\mathcal{P}(c_{\mu})\vk$
and $\vk_{1}=\vk$.

For a certain MLG $M_{\vk},$ not all its irreducible coreps are useful.
Only those compatible with the ireps of the translation group are
important in physics. They, denoted as $D_{\vk}^{p}\ (p=1,2,\cdots),$
satisfy the following relation
\begin{equation}
D_{\vk}^{p}(\{\alpha|\bm{v}+\bm{t}\})=e^{-i\vk\cdot\bm{t}}D_{\vk}^{p}(\{\alpha|\bm{v}\})\ \ \text{for }\forall\bm{t}\in\mathbb{L},\ \ ,\label{eq:v+t}
\end{equation}
in which $\alpha$ can be $R$ or $\mathcal{T}R$ and $\mathbb{L}$
is the set of all lattice vectors. These coreps are called ``allowed''
irreducible coreps of $M_{\vk}$ and are also termed ``small coreps''.
As for the own irreducible coreps of the MSG $M$, they can be obtained
through the induction from the small coreps. That is to say, $D_{\vk}^{p}\Uparrow M$
is an irreducible corep of $M$. For each $\vk_{\mu}$ in the magnetic
star $^{*}\vk$, all coreps $D_{\vk_{\mu}}^{p}\Uparrow M\,(\mu=1,\cdots,r)$
are equivalent to each other. Therefore, the MSG corep $D_{\vk}^{p}\Uparrow M$
is said to belong to the whole magnetic star $^{*}\vk$. It's worth
reminding that two different concepts are involved here: ``MLG corep''
such as $D_{\vk}^{p}$ and ``MSG corep'' such as $D_{\vk}^{p}\Uparrow M$.
Despite this, when ``corep of/for MSG'' is generally mentioned,
its meaning may cover both MLG corep and MSG corep, because the MLG
corep $D_{\vk}^{p}$ of $M_{\vk}$ can also be said to be the small
corep of the MSG $M$ (at $\vk$). To differentiate from ``small
corep'' (e.g. $D_{\vk}^{p}$) explicitly, MSG corep (e.g. $D_{\vk}^{p}\Uparrow M$)
can also be called the ``full corep'' of MSG. 

The key to the corep theory of MSG is to obtain the small coreps of
MLG. Suppose $G$ is the maximal unitary subgroup of $M$. Then $G$
has to be one of the 230 SGs. And the LG of $G$ at $\vk$, denoted
as $G_{\vk},$ is also the maximal unitary subgroup of the MLG $M_{\vk}$.
If $M_{\vk}$ is unitary, then $M_{\vk}=G_{\vk}$ and the small coreps
of $M_{\vk}$ are just the small reps of $G_{\vk}$. But if $M_{\vk}$
is nonunitary, the small coreps of $M_{\vk}$ can be calculated from
the small reps of $G_{\vk}$ via Eqs. (\ref{eq:corepa})\textendash (\ref{eq:corepc}).
In either case, all the small reps of $G_{\vk}$ are known thanks
to the \textsf{SpaceGroupIrep} package, and the small coreps of $M_{\vk}$
can be obtained subsequently.

For the full coreps of nonunitary MSGs (type-II, -III, and -IV), we
extend the three cases (a), (b), and (c) to four types (a), (b), (c),
and (x). Both (a) and (b) keep unchanged. But the case (c) is further
divided into types (c) and (x). Suppose that $\Gamma_{\vk}^{q}$ is
a small rep of $G_{\vk}$ and the small corep of $M_{\vk}$ derived
from $\Gamma_{\vk}^{q}$ via Eqs. (\ref{eq:corepa})\textendash (\ref{eq:corepc})
is $D_{\vk}^{p}=\mathcal{D}\Gamma_{\vk}^{q}.$ Then the type of the
full corep $D_{\vk}^{p}\Uparrow M$ can be calculated via 
\begin{equation}
\sum_{c\in(M_{\vk}-G_{\vk})\cosets T}\chi_{\vk}^{q}(c^{2})=\begin{cases}
+|G_{\vk}\cosets T| & \text{: type (a), case (a)}\\
-|G_{\vk}\cosets T| & \text{: type (b), case (b)}\\
0\ (M_{\vk}\ne G_{\vk}) & \text{: type (c), case (c)}\\
0\ (M_{\vk}=G_{\vk}) & \text{: type (x), case (c)}
\end{cases}\label{eq:type}
\end{equation}
where $(M_{\vk}-G_{\vk})\cosets T$ means the set of coset representatives
for the cosets of the translation group $T$ which are contained in
$M_{\vk}-G_{\vk}$, and $\chi_{\vk}^{q}$ is the character of $\Gamma_{\vk}^{q}.$
Although the above types are defined for the full corep $D_{\vk}^{p}\Uparrow M$,
they are also meaningful for the small corep $D_{\vk}^{p}.$ The type
(x) in Eq. (\ref{eq:type}) implies that $M_{\vk}$ can be unitary
even if $M$ is nonunitary, and for this type additional degeneracy
does not exist at the $\vk$ point but does appear in the whole Brillouin
zone (BZ) when the symmetry group changes from $G$ to $M$. According
to Eq. (\ref{eq:type}), all the small coreps and full coreps of type-I
MSGs belong to type (x). Here, we conclude that the four types are
defined for both unitary and nonunitary groups but the three cases
are only defined for nonunitary groups. The possible types and cases,
together with their relations to the additional degeneracy, are listed
in Table \ref{tab:types}. 

\begin{table}

\caption{The possible types and cases of both the small corep $D_{\protect\vk}^{p}$
and the full corep $D_{\protect\vk}^{p}\Uparrow M$ for the MSG $M$.
The symbol \textendash{} means undefined. The last two columns show
whether the additional degeneracy will appear when the symmetry group
changes from $G$ to $M$ in which $G$ is the maximal unitary subgroup
of $M$.\label{tab:types}}

\begin{centering}
\begin{tabular}{clccccccc}
\hline 
\multicolumn{2}{c}{} & \multicolumn{2}{c}{$D_{\vk}^{p}$} &  & \multicolumn{2}{c}{$D_{\vk}^{p}\Uparrow M$} & doubled & doubled \tabularnewline
\cline{3-4} \cline{4-4} \cline{6-7} \cline{7-7} 
\multicolumn{2}{c}{\raisebox{0.5em}[0.5em]{MSG type}} & type & case &  & type & case & at $\vk$ & in BZ\tabularnewline
\hline 
\multirow{4}{*}{%
\parbox[c]{7ex}{%
\begin{center}
non-\\
unitary
\end{center}%
}
} & \multirow{4}{*}{%
\parbox[t]{8ex}{%
type-II\\
type-III\\
type-IV
}
} & a & a &  & a & a & no & no\tabularnewline
 &  & b & b &  & b & b & yes & no\tabularnewline
 &  & c & c &  & c & c & yes & no\tabularnewline
 &  & x & \textendash{} &  & x & c & no & yes\tabularnewline
unitary & type-I & x & \textendash{} &  & x & \textendash{} & no & no\tabularnewline
\hline 
\end{tabular}
\par\end{centering}
\end{table}

\section{Files and installation}

The \textsf{MSGCorep} package is developed in Wolfram language and
can be used in the Mathematica software with version $\ge11.2$. This
package is dependent on the \textsf{SpaceGroupIrep} package, so users
have to make sure \textsf{SpaceGroupIrep} (at best the latest version)
properly installed prior to the installation of \textsf{MSGCorep}.
The \textsf{MSGCorep} package mainly contains the files \textsf{MSGCorep.wl},
\textsf{MSGData.wl}, \textsf{Usage.wl}, and \textsf{libMLGCorep.mx}.
The first three are core files and the fourth one contains the data
of MLG coreps which are used only by the \lstinline!getBandCorep!
function. We also supply a file \textsf{libMLGCorep.mx\_RaspberryPi}
as an alternative version of \textsf{libMLGCorep.mx} working for raspberry
pi platform, because Mathematica can be used free of charge on raspberry
pi for non-commercial usage \citep{rasppi}. To install the package,
just put the \textsf{MSGCorep} folder containing the above-mentioned
files to the path \lstinline!$UserBaseDirectory/Applications/! or
other available paths \citep{Liu_Yao_2021_265_107993__SpaceGroupIrep}.
After installation, users can use the following statement to load
the \textsf{MSGCorep} package and start using it.

\begin{lstlisting}[backgroundcolor={\color{yellow!5!white}},mathescape=true,literate={`}{\textasciigrave}{1}]
<<"MSGCorep`"   (* or just: <<MSGCorep` *)
\end{lstlisting}
Here we provide several useful tips for users {[}cf. Sec. 1 in the
supplementary material (SM){]}:
\begin{itemize}
\item Use \lstinline[literate={`}{\textasciigrave}{1}]!?MSGCorep`*! to
browse all available functions in the package or \lstinline[literate={`}{\textasciigrave}{1}]!?MSGCorep`*MLG*!
for all functions containing \lstinline[literate={`}{\textasciigrave}{1}]!MLG!
in their names such as \lstinline[literate={`}{\textasciigrave}{1}]!getMLGElem!.
\item Use \lstinline!?getBandCorep! to see the information about a specific
function such as \lstinline!getBandCorep!.
\item Many functions have options which can enrich their functionalities.
Use \lstinline[mathescape=true]!Options[$\ensuremath{\textit{\rmfamily fun}}$]!
to see the available options and their default values of a function
\textit{fun}.
\end{itemize}

\section{Conventions}

In both \textsf{SpaceGroupIrep} and \textsf{MSGCorep} packages, we
follow the convention used in the classic book written by C. J. Bradley
and A. P. Cracknell (the BC book) \citep{BCbook} and call it ``BC
convention''. This convention is composed of various definitions
and tables in the BC book, such as the basic vectors of primitive
cells (defined in the Tab. 3.1 in the BC book, hereafter referred
to as ``BC-Tab. 3.1''), the names and operations of rotations (BC-Tabs.
1.4, 3.2, 3.4, and 6.7{*}), the elements of SGs (BC-Tab. 3.7{*}) and
MSGs (BC-Tabs. 7.2{*} and 7.3{*}), the high-symmetry k-points (BC-Tab.
3.6{*}), small reps (BC-Tabs. 5.7{*}, 5.11, 6.13{*}, and 6.15) and
their labels (BC-Tabs. 5.8{*} and 6.14{*}). Note that the tables with
{*} have corrections or adaptions given in this paper or the SM of
Ref. \citep{Liu_Yao_2021_265_107993__SpaceGroupIrep}.

As is well known, there are mainly two types of MSG notations (including
numbers and symbols) which have been widely used for tens of years,
namely the Belov-Neronova-Smirnova (BNS) notations \citep{Belov_Smirnova_1957_1_487__1651,Belov_Smirnova_1957_2_311__Shubnikov}
and the Opechowski-Guccione (OG) notations \citep{OG1965}. Here we
use the BNS MSG notations listed in BC-Tab. 7.4{*} with some corrections
and adaptions (details are given in \ref{sec:fixTab} and Table \ref{tab:fixTab7.4}).
We made the adaptions to keep the BNS symbols the same as those in
the latest and systematic MSG monograph written by Litvin in 2013
\citep{Litvin2013MGT}. We regard these BNS symbols as the standard
MSG symbols. It should be pointed out that the orientation of conventional
cell in BC convention, shortened as BC orientation, is not always
the same as the default orientation in ``International Tables for
Crystallography, Volume A'' (hereafter referred to as ITA \citep{ITA}),
shortened as ITA orientation. The differences exist mainly in monoclinic
and orthorhombic crystal systems. In principle, a self-consistent
MSG symbol should reflect the cell orientation. But note that the
standard BNS symbols that we use are not always self-consistent, because
they correspond to the ITA orientation but not always to the BC orientation.
For comparison purpose, we list all the symbols in BC orientation
in Table S1 in the SM if they are not the same with the standard symbols
(run \lstinline!showMSGSym[]! to get the table). Besides, we also
suppply the OG numbers and symbols, which are consistent with those
given by Litvin \citep{Litvin2013MGT}. It's worth noting that, there
are some differences between the symbols used here (also by Litvin)
and those in ISO-MAG \citep{iso-mag} and BCS, see the details in
\ref{sec:diffSym} and Tables \ref{tab:diffsymOG} and \ref{tab:diffsymBNS}.

\begin{figure}
\begin{centering}
\includegraphics[width=16cm]{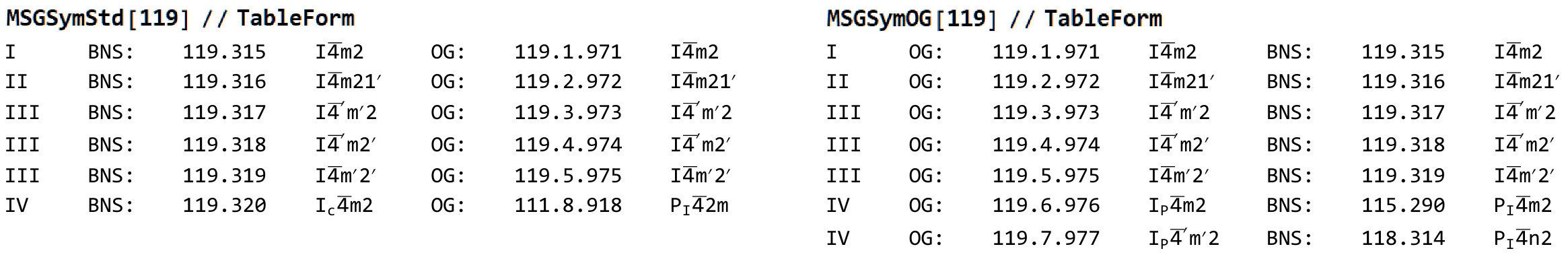}
\par\end{centering}
\caption{\label{fig:msgsym}Examples of \lstinline!MSGSymStd! and \lstinline!MSGSymOG!
if their arguments are only the family number of MSG. }

\end{figure}

Users can use the following functions to get the MSG symbols,

\begin{lstlisting}[backgroundcolor={\color{yellow!5!white}},mathescape=true]
MSGSymStd[{28,92}]    (* gives standard BNS symbol of 28.92: $\color{green!30!gray}P_ama2$ *)
MSGSymBC[{28,92}]     (* gives the BNS symbol in BC orientation: $\color{green!30!gray}P_bbm2$ *)
MSGSymOG[{25,12,166}] (* gives the OG symbol of 25.12.166: $\color{green!30!gray}P_{2a}m'm'2$ *)
\end{lstlisting}
in which we use a list of two (three) integers \lstinline!{28,92}!
(\lstinline!{25,12,166}!) to describe the BNS (OG) MSG number 28.92
(25.12.166). The OG number \lstinline!{25,12,166}! can be obtained
by \lstinline!BNStoOG[{28,92}]!, and reversely \lstinline!OGtoBNS[{25,12,166}]!
returns \lstinline!{28,92}!. If only the family number of an MSG
is given as the argument, \lstinline!MSGSymStd! (\lstinline!MSGSymOG!)
will return all possible MSGs in this family. For example, \lstinline!MSGSymStd[119]!
(\lstinline!MSGSymOG[119]!) gives all the MSG symbols whose BNS (OG)
numbers begin with 119, together with the correspondence to the OG
(BNS) notations, as shown in the left (right) panel of Fig. \ref{fig:msgsym}.
From Fig. \ref{fig:msgsym} one can see that the BNS family of MSGs
are different from the OG family of MSGs with the same family number.
In addition, one can also use \lstinline!showMSGSym[119]! to show
the symbols of BNS family 119 in a formatted table form, or use \lstinline!showMSGSym[110;;120]!
to show a range of families from 110 to 120.

When we construct type-III MSGs according to BC-Tab. 7.2, we find
that the ``colored'' generating elements given in BC-Tab. 7.2 are
mistaken for many MSGs, mainly involving orthorhombic and hexagonal
systems. Here, a colored element, also called a primed element, means
that it's an anti-unitary element and the multiplied $\tr$ is usually
represented by a $'$. For example, after being colored, $\{R|\bm{v}\}$
becomes $\{R|\bm{v}\}'$ which is equivalent to $\{R|\bm{v}\}\tr$,
and $\{R|\bm{v}\}'$ is usually shown in red color as ${\color{red}\{R|\bm{v}\}'}$
to attract attention. Taking the orthorhombic MSG 28.89 ($Pm'a2')$
for example, the $\sigma_{x}$ given in BC-Tab. 7.2 means that the
generating elements of $Pm'a2'$ can be obtained by first getting
the generating elements of SG 28 ($Pma2)$, i.e. $\{\sigma_{x}|\frac{1}{2}00\}$
and $\{\sigma_{y}|000\}$, from BC-Tab. 3.7 (or run \lstinline!SGGenElem[[28,1]]!
to get them) and then making the specified one, i.e. $\{\sigma_{x}|\frac{1}{2}00\}$
which is specified by its rotation part $\sigma_{x}$, colored. However,
thus obtained generating elements, i.e. ${\color{red}\{\sigma_{x}|\frac{1}{2}00\}'}$
and $\{\sigma_{y}|000\}$, are not consistent with the symbol $Pm'a2'$.
The symbol $Pm'a2'$ tells us that it has a primed mirror (i.e. $m'$)
and an unprimed glide plane (i.e. $a$), but ${\color{red}\{\sigma_{x}|\frac{1}{2}00\}'}$
and $\{\sigma_{y}|000\}$ are a primed glide reflection and an unprimed
mirror reflection respectively. More specifically, ${\color{red}\{\sigma_{x}|\frac{1}{2}00\}'}$
($\{\sigma_{y}|000\}$) gives a primed glide plane $b'$\footnote{Note that in BC convention, the basic vectors of orthorhombic primitive
lattice are $\bm{t}_{1}=(0,-b,0)=-\bm{b}$, $\bm{t}_{2}=(a,0,0)=\bm{a}$,
and $\bm{t}_{3}=(0,0,c)=\bm{c}$, and the translation part $(\frac{1}{2}00)$
in $\{\sigma_{x}|\frac{1}{2}00\}$ represents the translation vector
$\frac{1}{2}\bm{t}_{1}=-\frac{1}{2}\bm{b}$. Therefore, $\{\sigma_{x}|\frac{1}{2}00\}$
is a glide operation with glide direction $\bm{b}$ and it should
be denoted as $b$ in the SG symbol. Furthermore, ${\color{red}\{\sigma_{x}|\frac{1}{2}00\}'}$
gives a primed glide plane $b'$ in the MSG symbol..} (unprimed mirror $m$) with normal direction along $\bm{x}$ ($\bm{y}$),
and it should correspond to the first (second) place holder $\Square$
in the MSG symbol $P\Square\Square2'$. So, the MSG with generating
elements ${\color{red}\{\sigma_{x}|\frac{1}{2}00\}'}$ and $\{\sigma_{y}|000\}$
should be $Pb'm2'$. Recalling that in BC convention the actual orientation
of SG 28 is $\bm{b\bar{a}c}$ (refer to the Table 5 in \citep{Liu_Yao_2021_265_107993__SpaceGroupIrep}
or run \lstinline!BCOrientation[28]! to get it), the symbol $Pb'm2'$
is actually in BC orientation, and becomes $Pma'2'$ after converting
to ITA orientation (i.e. $\bm{abc}$). However, $Pma'2'$ is MSG 28.90,
not 29.89! If the MSG generating elements are $\{\sigma_{x}|\frac{1}{2}00\}$
and ${\color{red}\{\sigma_{y}|000\}'}$, similar analyses directly
give the MSG symbol $Pbm'2'$, corresponding to $Pm'a2'$ (MSG 28.89)
in ITA orientation. To this point, one can see that BC-Tab. 7.2 gives
mistaken colored generating elements for MSGs 28.89 and 28.90. We
think the reason of this mistake may be that Bradley and Cracknell
simply copied the data in the original BNS table in \citep{Belov_Smirnova_1957_2_311__Shubnikov}
without essential adaptions to make them consistent with BC-Tab. 3.7.
In Table \ref{tab:fixTab7.2} we list the corrections to all the mistaken
colored generating elements in BC-Tab. 7.2. And in the Table S2 of
SM elements for each MSG are listed.

When we construct type-IV MSGs, we find that BC-Tab. 7.3 should also
be adapted. Firstly, BC-Tab. 7.3 is not complete, lacking six black-white
lattices (see Table \ref{tab:BWlatt} for details), such as the $P_{c}$
black-white lattice of MSG 7.28 ($P_{c}C$). Secondly, for all MSGs
of orthorhombic crystal system, we should use the BNS symbols in BC
orientation to obtain their black-white lattices. For example, MSG
51.300 should be constructed through the black-white lattice $P_{a}$
in its symbol in BC orientation, namely $P_{a}cmm$, not the $P_{c}$
in the standard symbol $P_{c}mma$. Only thus can we obtain the correct
elements of $M_{51.300}$ (MSG 51.300) based on $G_{51}$ (SG 51)
whose elements are consistent with BC-Tab. 3.7, and the equation is
$M_{51.300}=G_{51}+\tr\{E|\bm{t}_{0}\}G_{51}$ with $\bm{t}_{0}$
taking the value $\frac{1}{2}\bm{t}_{2}$ of the orthorhombic $P_{a}$
in BC-Tab. 7.3. Accordingly, the $A_{a}$, $A_{c}$, and $A_{C}$
in BC-Tab. 7.3 are of no use because the actual black-white lattices
are $C_{c}$, $C_{a}$, and $C_{A}$ respectively for corresponding
symbols in BC orientation\footnote{Taking MSG 40.208 for example, one can verify this by running \lstinline!MSGSymStd[\{40,208\}]!
and \lstinline!MSGSymBC[\{40,208\}]! which returns $A_{a}ma2$ and
$C_{c}c2m$ respectively. }. 

In short, in order to construct MSGs from BC-Tabs. 7.2 and 7.3 correctly,
the actual BC orientation has to be considered and the two tables
have to be corrected or adapted accordingly, or else the obtained
MSG elements may not match the MSG numbers and symbols. The SM of
Ref. \citep{Tang_Wan_2021_104_85137_2103.08477_Exhaustive} mentions
that there are 126 MSGs whose numbers in the BC book are different
from those on BCS and it also gives the correspondence between them.
In fact, those 126 MSGs are merely constructed from the original BC-Tabs.
7.2 and 7.3 literally in Ref. \citep{Tang_Wan_2021_104_85137_2103.08477_Exhaustive},
which leads to the seeming differences between the BC book and BCS.
However, the BNS numbers in both the BC book and on BCS should not
be different from the original definitions by Belov, Neronova, and
Smirnova, even if the BNS symbols may be different according to the
cell orientation.

\section{Functionalities of \textsf{MSGCorep}}

\subsection{Group elements and multiplication }

In \textsf{SpaceGroupIrep} an SG element $\{R|\bm{v}\}$ is described
as \lstinline!{Rname,{v1,v2,v3}}! in the code, while in \textsf{MSGCorep}
an MSG element is described as \lstinline!{Rname,{v1,v2,v3},au}!
in which \lstinline!au! describes whether this element is anti-unitary
whose value can only be 0 (no) or 1 (yes). For example, as MSG elements,
$\{C_{2z}|00\frac{1}{2}\}$ and $\{C_{2z}|00\frac{1}{2}\}'$ are described
as \lstinline!{"C2z",{0,0,1/2},0}! and \lstinline!{"C2z",{0,0,1/2},1}!
respectively in the code. To get the list of the elements in an MSG
with BNS number $n.m$, one can use \lstinline!getMSGElem[{n,m}]!,
and similarly one can use \lstinline!getMLGElem[{n,m},k]! to get
the elements in the MLG of \lstinline!k! in which \lstinline!k!
can be either the name string of a BC standard k-point\footnote{A ``BC standard k-point'', denoted as $\vk_{{\rm BC}}$, is a k-point
whose coordinates are defined in BC-Tab. 3.6{*}.} or the fractional coordinates of any k-point. We have to point out
that what returns by \lstinline!getMSGElem! and \lstinline!getMLGElem!
are actually $M\cosets T$ and $M_{\vk}\cosets T$ respectively, namely
the coset representatives with respect to the translation group $T$.
If the option \lstinline!"double"->True! is used for the two functions,
the elements of the corresponding double MSG (or MLG) are returned.
For example,
\begin{lstlisting}[backgroundcolor={\color{yellow!5!white}},mathescape=true]
|In[1]:= |getMSGElem[{28,89}]
        getMLGElem[{28,89},"A"]  (* the coordinates of "A" is {0,u,1/2} *)
        getMLGElem[{28,89},"A","double"->True]
        showMSGSeitz/@%
|Out[1]=| [*{{"E",{0,0,0},0},{"C2z",{1/2,0,0},1},{"$\sigma$x",{1/2,0,0},0},{"$\sigma$y",{0,0,0},1}}*]
|Out[2]=| [*{{"E",{0,0,0},0},{"C2z",{1/2,0,0},1}}*]
|Out[3]=| [*{{"E",{0,0,0},0},{"barE",{0,0,0},0},{"C2z",{1/2,0,0},1},{"barC2z",{1/2,0,0},1}}*]
|Out[4]=| [*{$\scalebox{0.8}{\ensuremath{\{E|000\},
           \{\overline{E}|000\},\color{red}\{C_{2z}|\frac1200\}',\{\overline{C_{2z}}|\frac1200\}'}}$}*]
\end{lstlisting}
in which \lstinline!showMSGSeitz! shows an MSG element in the mathematical
form and highlights anti-unitary element in red color.

The multiplication of two MSG elements $\{R_{1}|\bm{v}\}^{\texttt{au1}\prime}$$\{R_{2}|\bm{w}\}^{\texttt{{au2}\ensuremath{\prime}}}=\{R_{1}R_{2}|R_{1}\bm{w}+\bm{v}\}^{(\texttt{au1}\oplus\texttt{au2})\prime}$,
the inversion of an MSG element $(\{R|\bm{v}\}^{\texttt{au}\prime})^{-1}=\{R^{-1}|-R^{-1}\bm{v}\}^{\texttt{au}\prime}$,
and the $n$-th power of an MSG element $(\{R|\bm{v}\}^{\texttt{au}\prime})^{n}$
can be calculated by \lstinline!MSGSeitzTimes!, \lstinline!MSGinvSeitz!,
and \lstinline!MSGpowerSeitz! in the code respectively. Here $\{R|\bm{v}\}^{\texttt{au}\prime}$
represents $\{R|\bm{v}\}$ if $\texttt{au}=0$ or $\{R|\bm{v}\}'$
if $\texttt{au}=1$, and the symbol $\oplus$ means ``exclusive or''.
All the three functions have versions for double MSGs with a \lstinline!DMSG!
prefix because double MSGs have different multiplication as follows:
\begin{equation}
\begin{cases}
\{R_{1}|\bm{v}\}\{R_{2}|\bm{w}\}=\{R_{1}R_{2}|R_{1}\bm{w}+\bm{v}\}\\
\{R_{1}|\bm{v}\}'\{R_{2}|\bm{w}\}=\{R_{1}|\bm{v}\}\{R_{2}|\bm{w}\}'=\{R_{1}R_{2}|R_{1}\bm{w}+\bm{v}\}' & \!\text{and }\\
\{R_{1}|\bm{v}\}'\{R_{2}|\bm{w}\}'=\{\bar{E}R_{1}R_{2}|R_{1}\bm{w}+\bm{v}\}
\end{cases}\begin{cases}
\{R|\bm{v}\}^{-1}=\{R^{-1}|-R^{-1}\bm{v}\}\\
(\{R|\bm{v}\}')^{-1}=\{\bar{E}R^{-1}|-R^{-1}\bm{v}\}'
\end{cases}\hspace{-1em}.
\end{equation}
Note the difference of $\tr^{2}$ between MSG and double MSG: for
MSG $(\{E|\bm{0}\}')^{2}=\{E|\bm{0}\}$ means $\tr^{2}=E$, while
for double MSG $(\{E|\bm{0}\}')^{2}=\{\bar{E}|\bm{0}\}$ means $\tr^{2}=\bar{E}$.
Finally the functions operating on (double) MSG elements are listed
below. 

\begin{lstlisting}[backgroundcolor={\color{yellow!5!white}},mathescape=true]
MSGSeitzTimes[brav][{R1,{v1,v2,v3},au1}, {R2,{w1,w2,w3},au2}, ...]
MSGinvSeitz[brav][{R,{v1,v2,v3},au}]
MSGpowerSeitz[brav][{R,{v1,v2,v3},au}, n]
DMSGSeitzTimes[brav][{R1,{v1,v2,v3},au1}, {R2,{w1,w2,w3},au2}, ...]
DMSGinvSeitz[brav][{R,{v1,v2,v3},au}]
DMSGpowerSeitz[brav][{R,{v1,v2,v3},au}, n]
\end{lstlisting}
In the above code, ``\lstinline!...!'' implies that \lstinline!MSGSeitzTimes!
and \lstinline!DMSGSeitzTimes! can calculate not only the multiplication
of two MSG elements, but also the continuous multiplication of three
or more MSG elements.

\subsection{Small Coreps}

\begin{figure}[th]
\begin{centering}
\includegraphics[width=14cm]{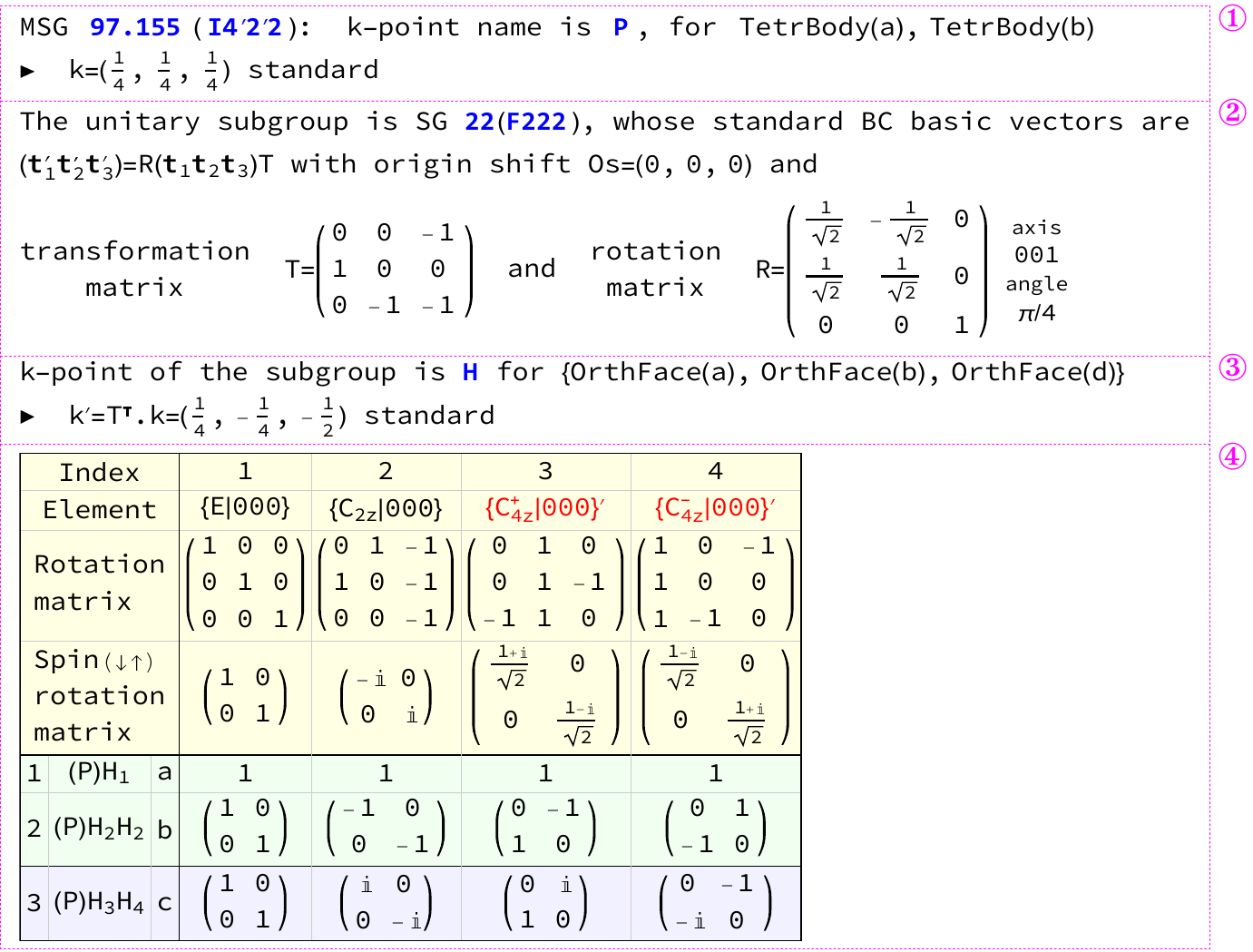}
\par\end{centering}
\caption{\label{fig:MLGCorep}The result of \lstinline!showMLGCorep[\{97,155\},"P"]!
which shows the small coreps of the $P$ point for MSG 97.155 $(I4'2'2)$.
Part \ding{172} shows the MSG notation, the k-point information such
as name and coordinates, and the types of BZs. Part \ding{173} shows
that SG 22 ($F222)$ is the maximal unitary subgroup of MSG 97.155
and it also shows the relations between the basic vectors of MSG 97.155
and the standard BC basic vectors of SG 22. Part \ding{174} shows
the converted k-point information from MSG 97.155 to SG 22. Parts
\ding{173} and \ding{174} exist only for type-III MSGs. Part \ding{175}
is the table which shows the small coreps. Light green backgroud shows
single-valued coreps and light blue background shows double-valued
coreps. The first, second, and third columns are the indexes, labels,
and types of the small coreps respectively.}

\end{figure}

The small coreps of any MLG can be obtained by \lstinline!getMLGCorep[{n,m},k]!
and shown in table form by \lstinline!showMLGCorep[{n,m},k]! in which
\lstinline!{n,m}! specifies the MSG number $n.m$ and \lstinline!k!
can be either the name string of a BC standard k-point or the numerical
fractional coordinates of any k-point. For using the corep data in
the code one can use \lstinline!getMLGCorep!, but if only for display
purpose one can just use \lstinline!showMLGCorep!. One example for
the small coreps of the $P$ point for MSG 97.155 $(I4'2'2)$ is shown
in Fig. \ref{fig:MLGCorep}, which is obtained by \lstinline!showMLGCorep[{97,155},"P"]!.
There are four parts \ding{172}--\ding{175} in Fig. \ref{fig:MLGCorep}.
Part \ding{172} shows the basic information, including the MSG number
97.155 and symbol $I4'2'2$, the k-point name $P$ and coordinates
$(\frac{1}{4}\frac{1}{4}\frac{1}{4})$, and the available types of
BZs \lstinline!TetrBody(a)! and \lstinline!TetrBody(b)!\footnote{Note that there are totally 22 types of BZs in BC convention and see
details in BC-Figs. 3.2\textendash 3.15 and the Table 1 in \citep{Liu_Yao_2021_265_107993__SpaceGroupIrep}. }. Strictly speaking, the k-point name shown here is only for the BC
standard k-point $\vk_{{\rm BC}}$, and for a type-II or -III k-point
(see the Sec. 2.3 in Ref. \citep{Liu_Yao_2021_265_107993__SpaceGroupIrep}
for details) this name is merely borrowed from $\vk_{{\rm BC}}$ to
name it. For example, if one calculate \lstinline!showMLGCorep[{97,155},{-1/4,-1/4,-1/4}]!,
the part \ding{172} of the result is shown below\\
\\[-0.4em]\centerline{\includegraphics[width=14cm]{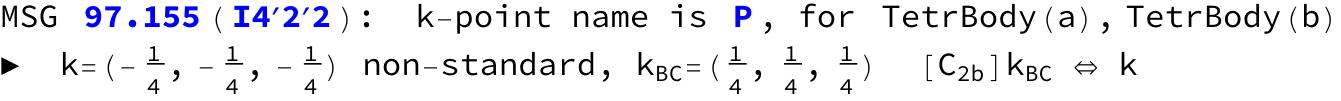}}\\
which implies that the input k-point $\vk=(\bar{\frac{1}{4}}\bar{\frac{1}{4}}\bar{\frac{1}{4}})$
is not standard and can borrow the name $P$ from $\vk_{{\rm BC}}=(\frac{1}{4}\frac{1}{4}\frac{1}{4})$
due to their relation $C_{2b}\vk_{{\rm BC}}\doteq\vk$. 

Parts \ding{173} and \ding{174} exist only for type-III MSGs. This
is because for type-I, -II, and -IV MSGs, the Bravais lattice of an
MSG $M$ is the same with that of its maximal unitary subgroup $G$
and their basic vectors are also the same, while for type-III MSGs,
both Bravais lattice and basic vectors may be different between $M$
and $G$. In order to utilize the existing irep data of SGs to construct
the coreps of MSGs, one has to find the relations between the primitive
cell of $M$ and the standard ``BC cell\footnote{Refer to the Sec. 8 in Ref. \citep{Liu_Yao_2021_265_107993__SpaceGroupIrep}
for more details.}'' of $G$ for type-III MSGs. In the example of Fig. \ref{fig:MLGCorep},
$M$ (MSG 97.155) has body-centered tetragonal Bravais lattice whose
basic vectors $\bm{t}_{1}$, $\bm{t}_{2}$, and $\bm{t}_{3}$ form
a basic-vector matrix 
\begin{equation}
[\bm{t}_{1}\ \bm{t}_{2}\ \bm{t}_{3}]=\begin{bmatrix}-a/2 & a/2 & a/2\\
a/2 & -a/2 & a/2\\
c/2 & c/2 & -c/2
\end{bmatrix}
\end{equation}
in which $a$ and $c$ are lattice constants, while the BC cell of
$G$ (SG 22) has face-centered orthorhombic Bravais lattice whose
basic vectors $\bm{t}_{1}'$, $\bm{t}_{2}'$, and $\bm{t}_{3}'$ are
\begin{equation}
[\bm{t}_{1}'\ \bm{t}_{2}'\ \bm{t}_{3}']=\begin{bmatrix}a'/2 & 0 & a'/2\\
0 & -b'/2 & -b'/2\\
c'/2 & c'/2 & 0
\end{bmatrix}
\end{equation}
in which $a'$, $b'$, and $c'$ are also lattice constants. Part
\ding{173} of Fig. \ref{fig:MLGCorep} shows the relationship between
the primitive cells of $M$ and $G$ in terms of transformation matrix
$\mathsf{\mathsf{T}}$, rotation matrix $\mathsf{R}$, and the origin
shift $\mathsf{Os}$ as follows 
\begin{equation}
\mathsf{R}[\bm{t}_{1}\ \bm{t}_{2}\ \bm{t}_{3}]\mathsf{T}=\begin{bmatrix}a/\sqrt{2} & 0 & a/\sqrt{2}\\
0 & -a/\sqrt{2} & -a/\sqrt{2}\\
c/\sqrt{2} & c/\sqrt{2} & 0
\end{bmatrix}=[\bm{t}_{1}'\ \bm{t}_{2}'\ \bm{t}_{3}']\ \ \ \ \Rightarrow\ \ \ \ a'=b'=\sqrt{2}a,\ c'=\sqrt{2}c.
\end{equation}
Here $\mathsf{Os}=(s_{1},s_{2},s_{3})$ gives the shift vector $s_{1}\bm{t}_{1}'+s_{2}\bm{t}_{2}'+s_{3}\bm{t}_{3}'$
from the origin of the cell of $G$ to the origin of the cell of $M$.
According to the relation $\vk'=\mathsf{T}^{T}\vk$, Part \ding{174}
of Fig. \ref{fig:MLGCorep} gives the k-point name $H$ and coordinates
$\vk'=(\frac{1}{4}\bar{\frac{1}{4}}\bar{\frac{1}{2}})$ of $G$ which
correspond to the input k-point $P$ of $M$ with coordinates $\vk=(\frac{1}{4}\frac{1}{4}\frac{1}{4})$. 

Part \ding{175} of Fig. \ref{fig:MLGCorep} is the table of small
coreps for the $P$ point of MSG 97.155. Both single-valued small
coreps (light green background) and double-valued small coreps (light
blue background) are shown, together with their labels (the second
column) and types (the third column). The row ``Element'' lists
the elements in the MLG. In fact only the coset representatives with
respect to $T$ are listed in the table and the elements whose rotation
parts have bars such as $\{\bar{E}|000\}$ are not shown because their
coreps can be simply obtained from the coreps of their counterparts
without bars. The simple relationship between them is 
\begin{equation}
\begin{cases}
D_{\vk}^{p}(\{\bar{R}|\bm{v}\}^{[\prime]})=D_{\vk}^{p}(\{R|\bm{v}\}^{[\prime]}) & \text{for single-valued coreps}\\
D_{\vk}^{p}(\{\bar{R}|\bm{v}\}^{[\prime]})=-D_{\vk}^{p}(\{R|\bm{v}\}^{[\prime]}) & \text{for double-valued coreps}
\end{cases}\label{eq:baredcorep}
\end{equation}
in which $^{[\prime]}$ means whether primed or not. And the coreps
of the elements whose translation parts differ from those given in
the table by a lattice vector can be obtained using Eq. (\ref{eq:v+t}).

The labels of small coreps are constructed from the labels of the
corresponding small reps from which the small coreps are derived.
\textbf{(i) Type-I, -II, and -IV MSGs.} A small corep, say $D$, of
type (a) or (x) is derived from only one small rep, say $\Gamma$,
and then $D$ just uses the label of $\Gamma$. For example, the small
corep $P_{3}$ of MSG 97.152 ($I4221')$ is of type (a) and it's derived
from the small rep $P_{3}$ of SG 97 ($I422$). One can check this
by running \lstinline!showMLGCorep[{97,152},"P"]! and \lstinline!showLGIrepTab[97,"P"]!.
If $D$ is of type (b), it contains two copies of $\Gamma$ for unitary
elements, and then the label of $D$ is two copies of the label of
$\Gamma$. For example, the small corep $A_{5}A_{5}$ of MSG 65.488
($C_{c}mmm)$ is of type (b) and it's derived from two copies of the
small rep $A_{5}$ of SG 65 ($mmm$). If $D$ is of type (c), it contains
two different small reps $\Gamma_{i}$ and $\Gamma_{j}$ for unitary
elements, and then the label of $D$ is the combination of the labels
of $\Gamma_{i}$ and $\Gamma_{j}$. For example, the small corep $P_{2}P_{4}$
of MSG 97.152 is of type (c) and it's the combination of the small
reps $P_{2}$ and $P_{4}$ of SG 97. \textbf{(ii) Type-III MSGs.}
First, an elementary label is constructed according to the above-mentioned
rules in case (i). Then an additional k-point name within parentheses
has to be prefixed to the elementary label to obtain the small corep
label. Taking the labels $(P)H_{1}$, $(P)H_{2}H_{2}$, and $(P)H_{3}H_{4}$
in Fig. \ref{fig:MLGCorep} for example, these labels contain the
information about two names $P$ and $H$ of the k-point: $P$ for
the MSG and $H$ for its maximal unitary subgroup. A k-point may have
different names for a type-III MSG and its maximal unitary subgroup,
and hence the information about both the k-point names is explicitly
given in the labels. For more information see the Appendix A of Ref.
\citep{Liu_Yao_2022_105_85117__Systematic}. In fact, the style of
labels for small coreps here is similar to that used by BCS. The difference
is that we always keep the parentheses prefix even if the k-point
names are the same for both the MSG and its maximal unitary subgroup,
such as the corep labels $(Z)Z_{1}$ and $(Z)Z_{2}Z_{4}$ for MSG
97.155.

\begin{figure}[t]
\begin{centering}
\includegraphics[width=14cm]{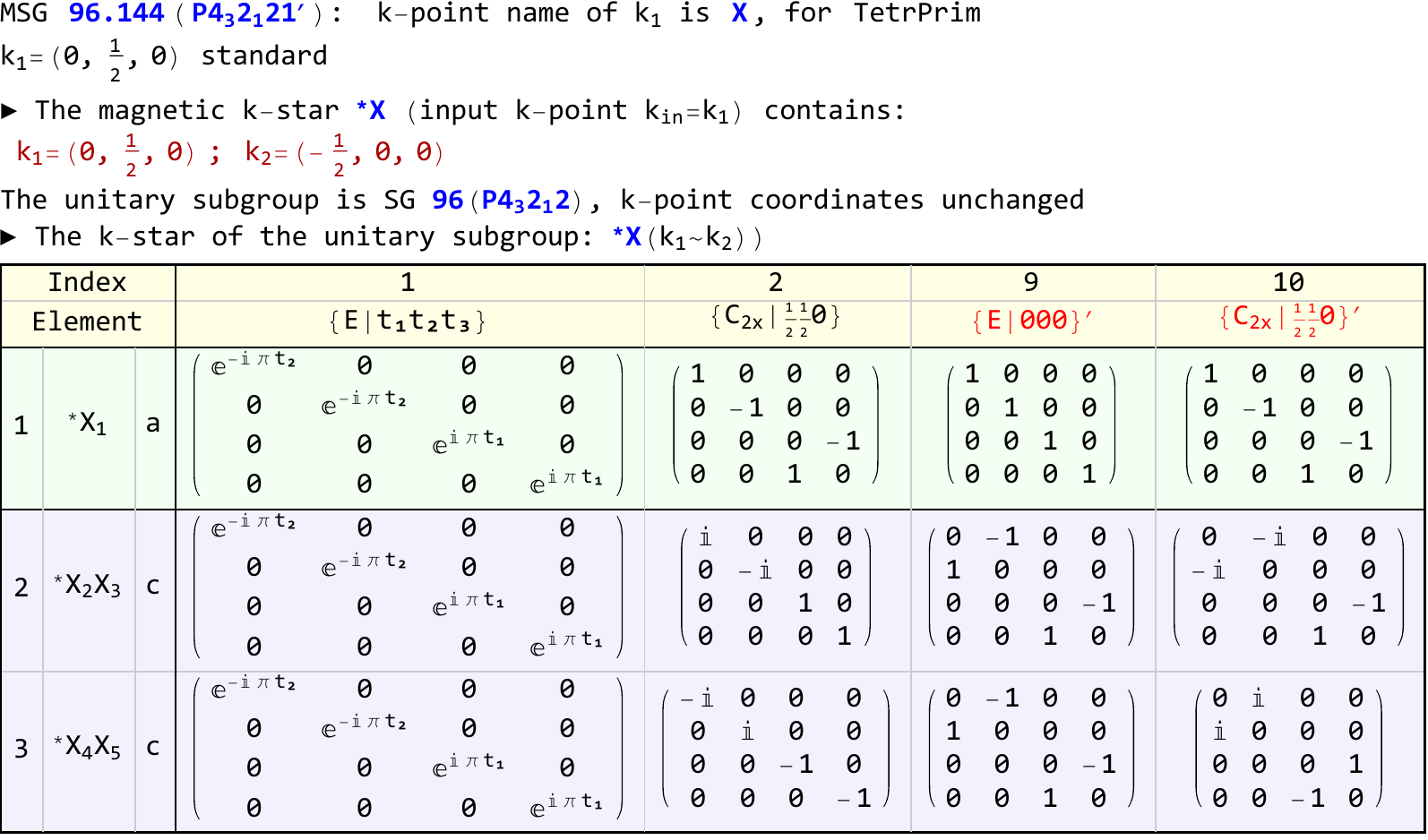}
\par\end{centering}
\caption{\label{fig:MSGCorep-1}The result of \lstinline!showMSGCorep[\{96,144\},"X","elem"->\{1,2,9,10\},"rotmat"->False]!
which shows the full coreps of the $^{*}\!X$ magnetic star for a
type-II MSG 96.144 $(P4_{3}2_{1}21')$. The coreps for only 4 elements
are shown due to the space limitation. Above the table, related information
is shown, including the notations of the MSG and its maximal unitary
subgroup, the name and coordinates of the magnetic k-star and the
input k-point, and so on. In the table of full coreps, light green
backgroud shows single-valued coreps and light blue background shows
double-valued coreps. The first, second, and third columns are the
indexes, labels, and types of the full coreps respectively.}
\end{figure}

\subsection{Full Coreps}

\begin{figure}[th]
\begin{centering}
\includegraphics[width=11cm]{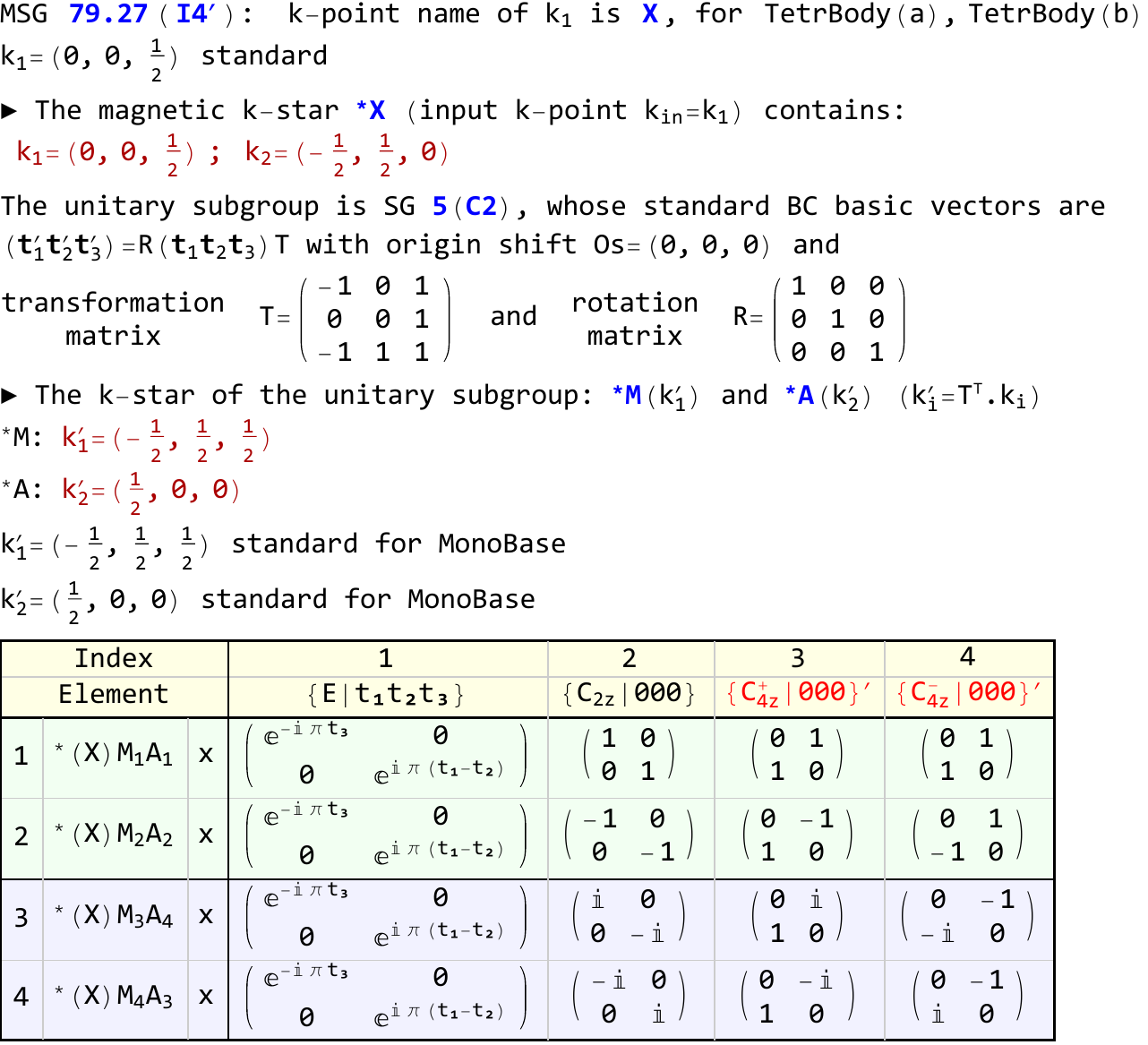}
\par\end{centering}
\caption{\label{fig:MSGCorep-2}The result of \lstinline!showMSGCorep[\{79,27\},"X","rotmat"->False]!
which shows the full coreps of the $^{*}\!X$ magnetic star for a
type-III MSG 79.27 $(I4')$. Above the table, related information
is shown, including the notations of the MSG and its maximal unitary
subgroup, the name and coordinates of the magnetic k-star and the
input k-point, the conversion relations between the MSG and its maximal
unitary subgroup similar to those in the parts \ding{173} and \ding{174}
of Fig. \ref{fig:MLGCorep}, and so on. In the table of full coreps,
light green backgroud shows single-valued coreps and light blue background
shows double-valued coreps. The first, second, and third columns are
the indexes, labels, and types of the full coreps respectively.}
\end{figure}

The full coreps of any MSG can be induced from the small coreps of
its MLG according to Eq. (\ref{eq:indcorep}). The calculated results
can be obtained by \lstinline!getMSGCorep[{n,m},k]! and shown in
table form by \lstinline!showMSGCorep[{n,m},k]! in which \lstinline!{n,m}!
specifies the MSG number $n.m$ and \lstinline!k! which is either
the name string or the numerical fractional coordinates of a k-point
specifies the magnetic star\footnote{The magnetic star of a k-point can be obtained by \lstinline!getMagKStar[\{n,m\},k]!.}
that includes it. Examples for the full coreps of the magnetic star
$^{*}\!X$ for MSG 96.144 $(P4_{3}2_{1}21')$ and MSG 79.27 $(I4')$
are shown in Fig. \ref{fig:MSGCorep-1} and Fig. \ref{fig:MSGCorep-2}
respectively. The result of \lstinline!showMSGCorep!, similar to
that of \lstinline!showMLGCorep!, is mainly composed of a text part
and a table part. The text part gives the information about the MSG,
the maximal unitary subgroup, the input k-point, the magnetic k-star,
and so on. By default, the table part gives the full coreps of the
elements in $T$ (see the $\{E|t_{1}t_{2}t_{3}\}$ column in the table
part) and in the coset representatives with respect to $T$ but not
including those whose rotation parts have bars. For the elements whose
rotation parts have bars, Eq. (\ref{eq:baredcorep}) is also available
if $D_{\vk}^{p}$ is substituted by a full corep . To save space,
rotation matrices are not shown in Figs. \ref{fig:MSGCorep-1} and
\ref{fig:MSGCorep-2} by using the option \lstinline!"rotmat"->False!,
and only coreps of four elements, i.e. the 1st, 2nd, 9th, and 10th
ones, are shown in Fig. \ref{fig:MSGCorep-1} by using the option
\lstinline!"elem"->{1,2,9,10}!.

In order to make the full coreps independent of which k-point is selected
from a given magnetic star as the input of \lstinline!showMSGCorep!,
the k-points in a given magnetic star have to be sorted in a fixed
order before the full coreps are constructed using Eq. (\ref{eq:indcorep}).
The input k-point of Fig. \ref{fig:MSGCorep-1} is \lstinline!"X"!
whose coordinates are $\vk_{{\rm in}}=(0\frac{1}{2}0)$. Its magnetic
star $^{*}\negmedspace X$ includes two k-points, namely $\vk_{1}=(0\frac{1}{2}0)$
and $\vk_{2}=(\bar{\frac{1}{2}}00)$. If one substitutes \lstinline!{-1/2,0,0}!
for \lstinline!"X"! as the input k-point, i.e. $\vk_{{\rm in}}=(\bar{\frac{1}{2}}00)$,
the obtained magnetic star keeps unchanged in the sense of equivalence
and the full coreps also keep the same\footnote{The result is not given here. One can run \lstinline!showMSGCorep[\{96,144\},\{-1/2,0,0\},"elem"->\{1,2,9,10\},"rotmat"->False]!
to check this. }. Note that one may find that the matrix elements of the coreps in
the $\{E|t_{1}t_{2}t_{3}\}$ column change seemingly from $e^{-i\pi t_{i}}$
to $e^{i\pi t_{i}}$ after the change of input k-point from \lstinline!"X"!
to \lstinline!{-1/2,0,0}!, but they are indeed equal because $t_{i}$
$(i=1,2,3)$ is always an integer here. 

The labels of full coreps are constructed based on the labels of the
small coreps that induce the full coreps and all the labels of full
coreps have an asterisk superscript on the left to indicate that full
coreps are defined according to magnetic k-stars. \textbf{(i)} For
full coreps of type (x) in type-I MSGs and full coreps of types (a),
(b), or (c) in all MSGs, the label of a full corep is constructed
by just putting an asterisk superscript on the left of the label of
the corresponding small corep. For example, the labels of small coreps
for the $X$ point of MSG 96.144 are $X_{1}$, $X_{2}X_{3}$, and
$X_{4}X_{5}$, and the labels of full coreps for the $^{*}\!X$ magnetic
star of MSG 96.144 are $^{*}\negmedspace X_{1}$, $^{*}\negmedspace X_{2}X_{3}$,
and $^{*}\negmedspace X_{4}X_{5}$ respectively. \textbf{(ii)} For
full coreps of type (x) in type-II, -III, and -IV MSGs, one magnetic
k-star of an MSG corresponds to two different k-stars of the maximal
unitary subgroup of the MSG. The label of such a full corep should
combine the labels of two small reps of the maximal unitary subgroup
from each of the two k-stars. To differentiate the two k-stars, two
different labels have to be used for them. There are three cases:
\textbf{1)} The two k-stars have different labels of BC convention.
For example, the $^{*}\negmedspace X$ magnetic star corresponds to
two different k-stars $^{*}\!M$ and $^{*}\negmedspace A$ of SG 5,
and hence its labels of full coreps are $^{*}(X)M_{1}A_{1}$, $^{*}(X)M_{2}A_{2}$,
$^{*}(X)M_{3}A_{4}$, and $^{*}(X)M_{4}A_{3}$, as shown in Fig. \ref{fig:MSGCorep-2}.
This case only occurs for type-III MSGs. \textbf{2)} The two k-stars
are negative of each other. In such a case, the label of the second
k-star is obtained by underlining the label of the first one. For
example, the $^{*}\!P$ magnetic star of MSG 44.230 corresponds to
the k-stars $^{*}\!P$ and $\underline{^{*}\!P}$ of SG 44, and its
full coreps have labels $^{*}\!P_{1}\underline{P_{1}}$, $^{*}\!P_{1}\underline{P_{2}}$,
$^{*}\!P_{3}\underline{P_{4}}$, and $^{*}\!P_{4}\underline{P_{3}}$.
All full coreps of type (x) for type-II and -IV MSGs are in this case,
but this case also occurs for type-III MSGs. \textbf{3)} The k-points
in both the k-stars can be identified as the same k-point name, but
the two k-stars are not negative of each other. In this case, one
k-star uses the BC k-point name, and the other k-star uses a k-point
name constructed by adding a letter ``A'' after the BC k-point name.
For example, the $^{*}\!W$ magnetic star of MSG 79.27 corresponds
to the k-stars $^{*}\negmedspace U$ and $^{*}\!U\!A$ of SG 5, and
its full coreps have labels $^{*}(W)U_{1}U\!A_{1}$, $^{*}(W)U_{2}U\!A_{2}$,
$^{*}(W)U_{3}U\!A_{4}$, and $^{*}(W)U_{4}U\!A_{3}$. This case also
occurs only for type-III MSGs.

\subsection{Direct product of full coreps}

\begin{figure}[t]
\begin{centering}
\includegraphics[width=13cm]{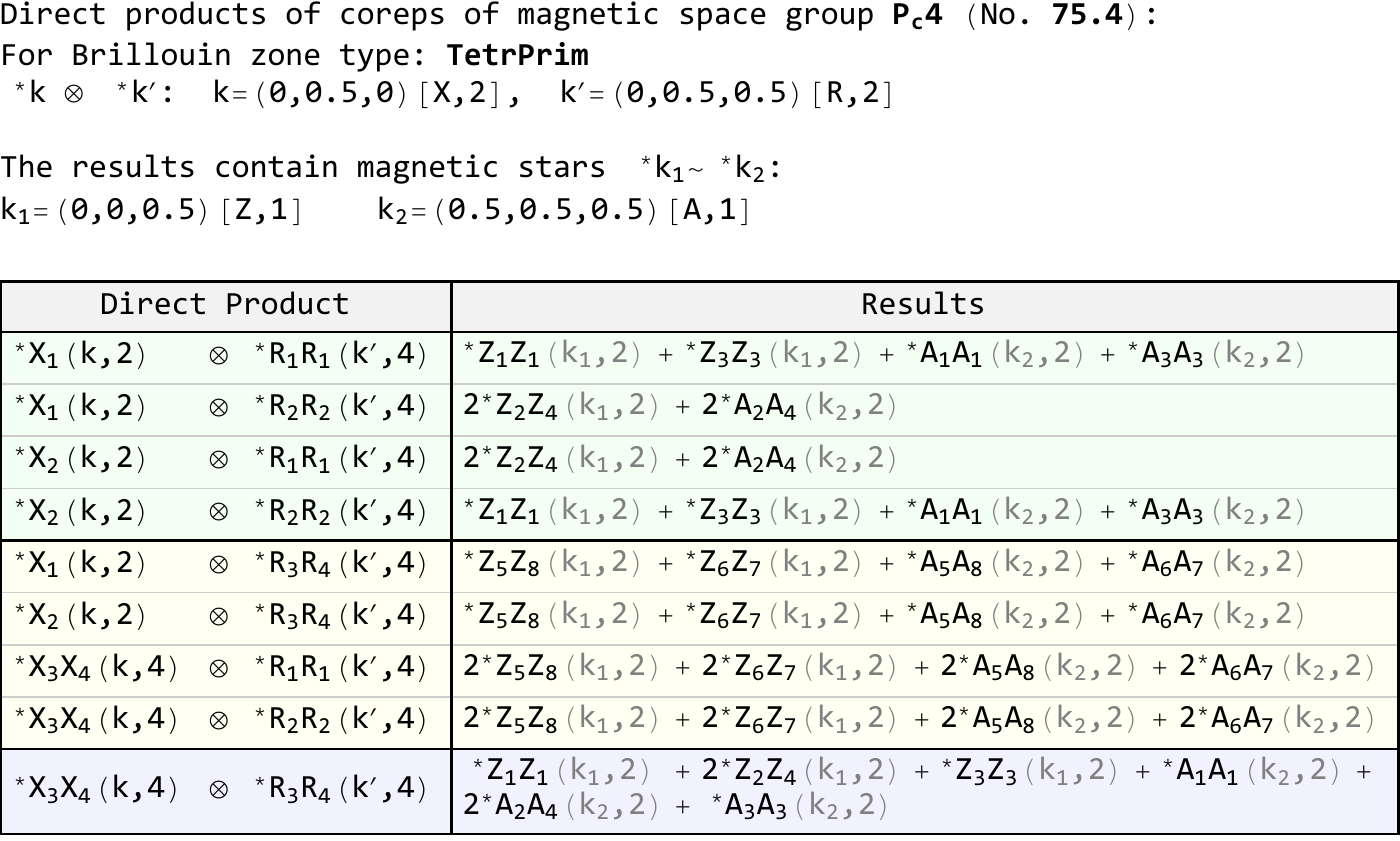}
\par\end{centering}
\caption{\label{fig:CorepDP}The result of \lstinline!showMSGCorepDirectProduct[\{75,4\},"X","R"]!
which shows the direct products of full coreps between two magnetic
stars $^{*}\!X$ and $^{*}\!R$ for MSG 75.4 $(P_{c}4)$. In the text
part above the table, the notation such as $[X,2]$ after a k-point
indicates the k-point name $(X)$ and the number of k-points (2) in
its magnetic star. The notation such as $(k_{1},2)$ in the table
indicates that the full corep in front of it, e.g. $^{*}\!Z_{2}Z_{4}$,
is for the magnetic star of $k_{1}$ and the dimension of this full
corep is 2. Light green (light blue) background stands for direct
products between two single-valued (double-valued) full coreps, and
light yellow background for direct products between a single-valued
full corep and a double-valued full corep. }
\end{figure}

The decomposition (or reduction) of the direct product (Kronecker
product) of two full coreps is useful to analyze the selection rules
of the quantum transition processes of electron, phonon, magnon and
so on in a crystal belonging to a certain MSG. Although the basic
calculation rules are all known, it's cumbersome and prone to error
to calculate the direct product manually. Accordingly, we have developd
the functions to calculate and show the decomposition of direct product
of full coreps for MSGs in the \textsf{MSGCorep} package, namely \lstinline!MSGCorepDirectProduct!
and \lstinline!showMSGCorepDirectProduct! respectively, because there
is no such a tool available to do this so far as we know. Suppose
$M$ is MSG $n.m$ and $G$ (an ordinary SG) is its maximal unitary
subgroup. $\Gamma^{i}$ and $\Gamma^{j}$ are two full ireps of $G$,
and $\mathcal{D}\Gamma_{i}$ and $\mathcal{D}\Gamma_{j}$ are the
corresponding two full coreps of $M$ derived from $\Gamma_{i}$ and
$\Gamma_{j}$ respectively. The direct product of $\mathcal{D}\Gamma_{i}$
and $\mathcal{D}\Gamma_{j}$ can be reduced to the direct sum of a
series of full coreps $\mathcal{D}\Gamma^{k}$ of $M$
\begin{equation}
\mathcal{D}\Gamma_{i}\otimes\mathcal{D}\Gamma_{j}=\bigoplus_{k}d_{ij,k}\mathcal{D}\Gamma^{k},
\end{equation}
and the task is to find the occurrence number (reduction coefficient)
$d_{ij,k}$ of the full corep $\mathcal{D}\Gamma^{k}$ in the decomposition
for each $k$. To calculate $d_{ij,k}$, one can first calculate the
reduction coefficient $c_{ij,k}$ in the direct product $\Gamma^{i}\otimes\Gamma^{j}=\bigoplus_{k}c_{ij,k}\Gamma^{k}$
for $G$, which can be done via the \lstinline!SGIrepDirectProduct!
function in the \textsf{SpaceGroupIrep} package. Then, $d_{ij,k}$'s
can be calculated through their relations to $c_{ij,k}$'s which are
given in BC-Tab. 7.8, and in the package one can use the following
codes to get the results.

\begin{lstlisting}[backgroundcolor={\color{yellow!5!white}},mathescape=true]
MSGCorepDirectProduct[{n,m}, k, kp]
showMSGCorepDirectProduct[{n,m}, k, kp]
\end{lstlisting}
In fact, the above codes calculate all the direct products between
two magnetic stars $^{*}\!\vk$ and $^{*}\!\vk'$ which are specified
via one k-point in each of them by \lstinline!k! and \lstinline!kp!
respectively. An example is shown in Fig. \ref{fig:CorepDP} for the
direct products between the full coreps of the magnetic stars $^{*}\negmedspace X$
and $^{*}\!R$ of MSG 75.4 ($P_{c}4$). There are three full coreps
for the magnetic star $^{*}\negmedspace X$, namely $^{*}\!X_{1}$,
$^{*}\!X_{2}$, and $^{*}\!X_{3}X_{4}$, and three full coreps for
$^{*}\!R$, namely $^{*}\!R_{1}R_{1}$, $^{*}\!R_{2}R_{2}$, and $^{*}\!R_{3}R_{4}$.
Fig. \ref{fig:CorepDP} shows all the nine direct products between
these two sets of full coreps, which are decomposed into the full
coreps of the magnetic stars $^{*}\negmedspace Z$ and $^{*}\!A$.

\subsection{Determine the small coreps of energy bands}

\begin{figure}[t]
\begin{centering}
\includegraphics[width=17.5cm]{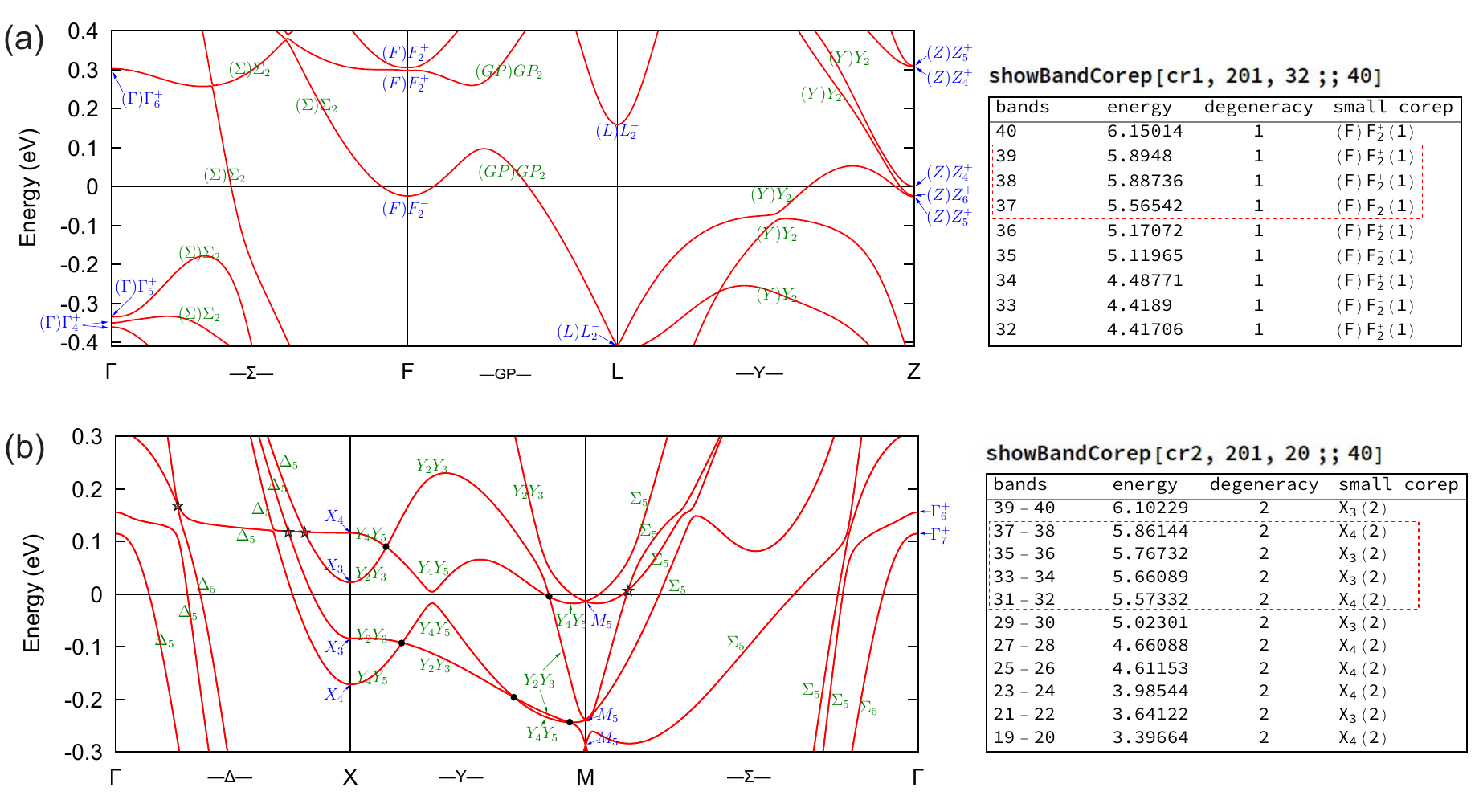}
\par\end{centering}
\caption{\label{fig:bandcorep}Examples of the bands and their small coreps
determined by \lstinline!getBandCorep! for (a) Mn$_{3}$NiN (MSG
166.101, $R\bar{3}m'$) and (b) 2Q phase of $\gamma$-Fe$_{x}$Mn$_{1-x}$
with $x=1$ (MSG 134.481, $P_{C}4_{2}/nnm$). The right panels of
(a) and (b) show the results of \lstinline!showBandCorep! which give
the small coreps at (a) $F$ and (b) $X$ k-points respectively. The
integer in the parentheses following the label of each small corep
is the dimension of the small corep. The dashed red boxes indicate
the states displayed in the band diagrams.}

\end{figure}

The small coreps of energy bands are very useful to construct the
$\vk\cdot\bm{p}$ models via symmetry systematically and to determine
the type of band crossings, just as done in Refs. \citep{Yu_Yao_2022_67_375_2102.01517_Encyclopedia,Liu_Yao_2022_105_85117__Systematic,Zhang_Yao_2022_105_104426_2112.10479_Encyclopedia,Tang_Wan_2021_104_85137_2103.08477_Exhaustive}.
These references directly tell one which small corep corresponds to
which $\vk\cdot\bm{p}$ model and what type of band crossing, but
they do not tell what is the small corep of a degenerate energy level
for a given band structure. To obtain the small coreps, two main steps
are needed: the first step is to calculate the characters of the elements
of the MLG at a desired k-point for each degenerate energy level,
and the second step is to determine small coreps by looking up the
character table of the MLG. In fact the characters of only unitary
elements need to be calculated because the equivalence of coreps only
need the characters of unitary elements. The first step can be done
by the tool \textsf{MagVasp2trace} \citep{MagVasp2trace,Xu_Bernevig_2020_586_702__High}
which is capable of post-processing the results of the first-principles
code VASP \citep{vasp1}, and the second step can be done by the \lstinline!readMagTrace!
and \lstinline!getBandCorep! functions in \textsf{MSGCorep}.

To successfully go through the procedure of determining small coreps,
there are three points needing attention. (1) The cell used for band
calculation has to conform to the BC convention, i.e. the rotation
matrices and translation components of the MSG elements determined
from the cell have to be compatible\footnote{Here ``compatible'' means the translation components can differ
by any integers.} with those from \lstinline!getMSGElem!. (2) The \textsf{msg.txt}
file which contains the MSG elements used by \textsf{MagVasp2trace}
also has to conform to the BC convention. Accordingly, do not use
the \textsf{msg.txt} file provided in the original \textsf{MagVasp2trace}
package, but use the files in the \textsf{BC\_MSG\_elements} directory
provided by \textsf{MSGCorep}. (3) For the VASP calculations which
turn on non-collinear spin but do not turn on spin-orbit coupling,
the wavefunctions are in fact spinors. In this case, \textsf{MagVasp2trace}
has to be modified a little as follows to output correct results.
\begin{lstlisting}[language=Fortran,showstringspaces=false,backgroundcolor={\color{yellow!5!white}},mathescape=true]
! read (nfst ,"( A90 )") chaps            ! line 111 of init.f90: comment it
  read (nfst ,"( A20L5 )") chtp15 ,FL (2) ! line 112 of init.f90: change A15 to A20
\end{lstlisting}

Then, one can first use \textsf{MagVasp2trace} to generate a \textsf{trace.txt}
file from the VASP outputs and then use the following functions in
\textsf{MSGCorep} to obtain the small coreps.
\begin{lstlisting}[backgroundcolor={\color{yellow!5!white}},mathescape=true]
(* put the trace.txt to the working directory (check this by Directory[]) *)
tr1=readMagTrace["trace.txt"];   (* read the trace.txt file from MagVasp2trace *)
cr1=getBandCorep[{n,m}, tr1];    (* determine the small coreps of the bands *)
showBandCorep[cr1, ik, ibs] (* show the small coreps of the ik-th k-point for bands ibs *)
\end{lstlisting}
Two examples are shown in Fig. \ref{fig:bandcorep}. Fig. \ref{fig:bandcorep}(a)
shows the bands and determined small coreps of Mn$_{3}$NiN with MSG
166.101 ($R\bar{3}m'$), a non-collinear antiferromagnetic material
with anomalous Hall effect, magneto-optical effect, and giant anomalous
Nernst effect \citep{Zhou_Mokrousov_2019_99_104428__Spin,Zhou_Yao_2020_4_24408__Giant}.
Another example is $\gamma$-Fe$_{x}$Mn$_{1-x}$ which is a topological
magneto-optical antiferromagnet when $0.4<x<0.8$ \citep{Feng_Yao_2020_11_118__Topological}.
For simplicity, here the 2Q phase for $x=1$ with MSG 134.481 ($P_{C}4_{2}/nnm$)
is taken for example and its bands and determined small coreps are
shown in Fig. \ref{fig:bandcorep}(b). The function \lstinline!showBandCorep!
can be used to extract the results for specified k-point and bands.
For example, the right panel of \ref{fig:bandcorep}(a) lists the
small coreps of the 201st k-point, which is the k-point $F$ in the
data of \lstinline!cr1!, for the bands from 32nd to 40th, and the
red dashed box indicates the three bands displayed in the band diagram,
namely the 37th, 38th, and 39th bands with small coreps $(F)F_{2}^{-}$,
$(F)F_{2}^{+}$, and $(F)F_{2}^{+}$ respectively. 

The identification of small coreps can help one to judge whether two
bands are crossed or anticrossed. For example, the positions marked
by \ding{73} along the $\Delta$ and $\Sigma$ paths in Fig. \ref{fig:bandcorep}(b)
are seemingly band crossings, but they are actually anticrossings
because the related bands have the same small coreps and the MSG symmetry
does not protect the crossings along these two paths. However, the
positions marked by $\bullet$ along the $Y$ path in Fig. \ref{fig:bandcorep}(b)
are true band crossings, because the related bands have different
small coreps. And one can also lookup the encyclopedia of emergent
particles for type-IV MSGs \citep{Zhang_Yao_2022_105_104426_2112.10479_Encyclopedia}
to find that each of these crossings along $Y$ path are in fact of
the type P-DNL, i.e. a point residing on a Dirac nodal line.

\subsection{Support for magnetic point group}

\begin{figure}[t]
\begin{centering}
\includegraphics[width=16cm]{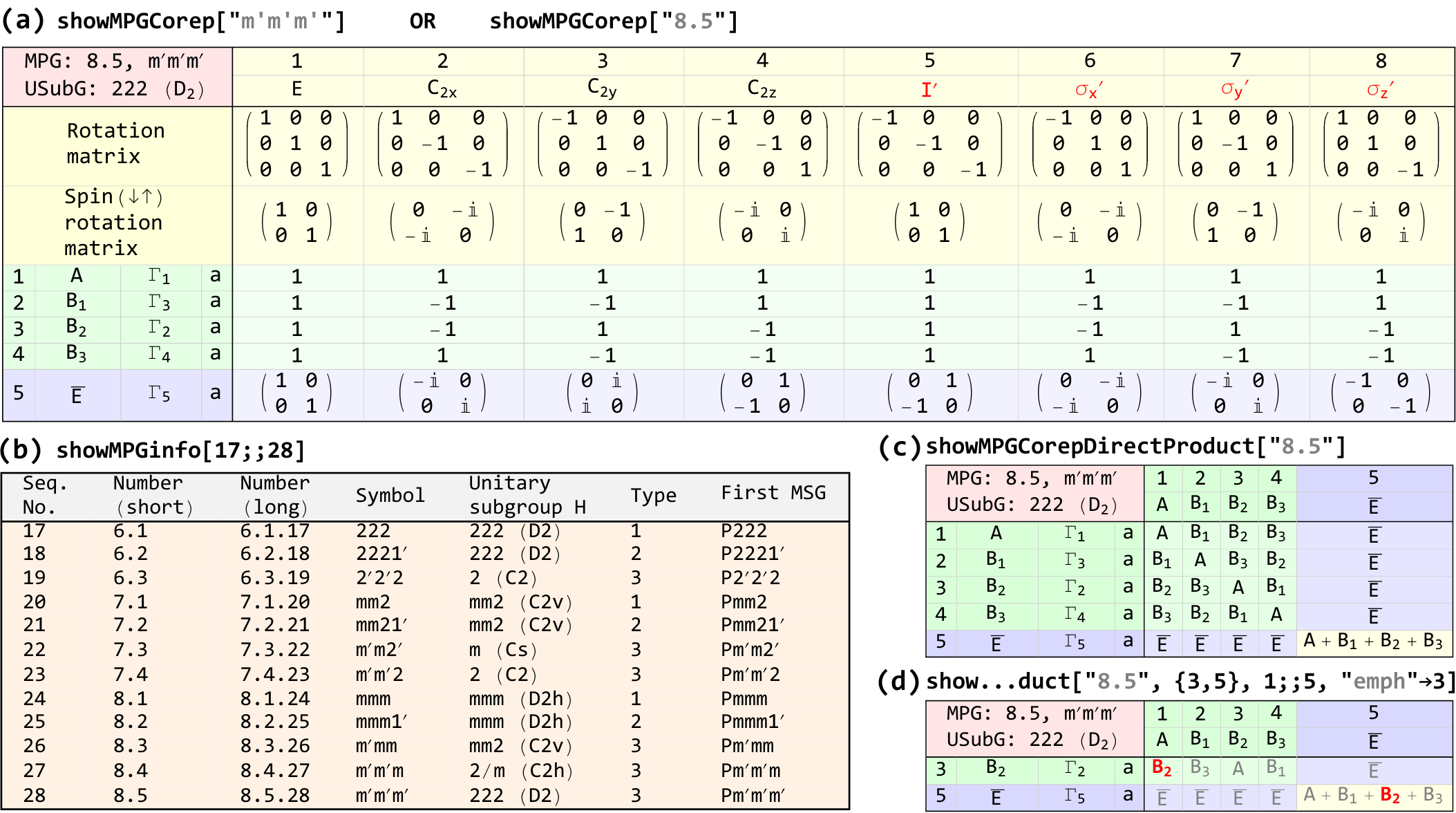}
\par\end{centering}
\caption{\label{fig:MPG}Examples of MPG-related functions. (a) The coreps
of the MPG 8.5 ($m'm'm'$) given by \lstinline!showMPGCorep!. Similar
to the result of \lstinline!showMLGCorep!, single-valued (double-valued)
coreps are shown in light green (light blue) background. The second
and third columns are the extended Mulliken labels and the $\Gamma$
labels of the coreps respectively, and the fourth column is the corep
type. (b) The information of the MPGs with sequence numbers from 17
to 28 given by \lstinline!showMPGinfo[17;;28]!. (c,d) \lstinline!showMPGCorepDirectProduct!
gives the direct products (c) between all the coreps or (d) between
two sets of specified coreps (\lstinline!\{3,5\}! and \lstinline!1;;5!)
with the third corep emphasized in red color (\lstinline!"emph"->3!). }

\end{figure}

As the finite subgroups of MSGs, magnetic point groups (MPGs) are
also supported by the \textsf{MSGCorep} package, including group multiplication,
coreps, and direct product of coreps for MPGs. There are in total
122 MPGs in three-dimensional space, including 32 type-I MPGs (i.e.
the 32 ordinary point groups), 32 type-II MPGs (also called gray point
groups), and 58 type-III MPGs (also called black-white point groups).
One can use \lstinline!showMPGinfo[]! to show the information of
the full list of the 122 MPGs, including their numbers, symbols, unitary
subgroups, types, etc. One can also specify the sequence numbers to
be shown. For example, \lstinline!showMPGinfo[17;;28]! gives the
MPGs with sequence numbers from 17 to 28, i.e. all the orthorhombic
MPGs, as shown in Fig. \ref{fig:MPG}(b). In order to simplify the
relationship between MPGs and MSGs, the symbol of an MPG is determined
by the BNS symbol of the first MSG whose point operations compose
the MPG and keep the BNS order of the MSG. For example, MSG 47.252
($Pm'm'm)$ is the first MSG with the MPG $m'm'm$ in BNS order, and
MPG 5.3 ($2'/m$) precedes MPG 5.4 ($2/m')$ because MSG 10.44 ($P2'/m$)
precedes MSG 10.45 ($P2/m')$. Accordingly, the MPG symbols and order
are not always the same with those in BC-Tab. 7.1, but they happen
to be the same with those used by BCS. 

An element of MPG is described as \lstinline!{Rname,au}! in the code
where \lstinline!Rname! is the name string of the rotation and \lstinline!au!
specifies the unitarity (0 for unitary and 1 for anti-unitary), such
as \lstinline!{"C2x",0}! for $C_{2x}$ and \lstinline!{"C2x",1}!
for ${\color{red}C_{2x}\vphantom{}'}(\equiv C_{2x}\tr)$. Similar
to \lstinline!getMSGElem!, one can use \lstinline!getMPGElem! to
obtain the elements of an MPG. For example, either \lstinline!getMPGElem["2'2'2"]!
or \lstinline!getMPGElem["6.3"]! gives the result of \lstinline[mathescape=true]!{{"E",0},{"C2z",0},{"C2x",1},{"C2y",1}}!.
The multiplication, inversion, and power of (double) MPG elements
can be calculated by the following functions.
\begin{lstlisting}[backgroundcolor={\color{yellow!5!white}},mathescape=true]
MagRotTimes[{R1,au1}, {R2,au2}]   (* the multiplication of two MPG elements *)
invMagRot[{R,au}]                 (* the inverse of an MPG element *)
powerMagRot[{R,au}, n]            (* the n-th power of an MPG element *)
DMagRotTimes[{R1,au1}, {R2,au2}]  (* the multiplication of two double MPG elements *)
invDMagRot[{R,au}]                (* the inverse of a double MPG element *)
powerDMagRot[{R,au}, n]           (* the n-th power of a double MPG element *)
\end{lstlisting}
In fact, as an argument of the above six functions, an MPG element
can also be expressed simply in a string form with a \lstinline!'!
representing $\tr$, such as \lstinline!"C2x'"! which is equivalent
to \lstinline!{"C2x",1}!. Furthermore, \lstinline!MagRotTimes! and
\lstinline!DMagRotTimes! can also calculate the continuous multiplication
of three MPG elements or more. For example, \lstinline!MagRotTimes["C2x", "C2a", "C31+'"]!
returns \lstinline!{"C4x+",1}!.

The coreps of an MPG can be obtained by \lstinline!getMPGCorep[mpg]!
and shown by \lstinline!showMPGCorep[mpg]! in which \lstinline!mpg!
is the symbol (or number) string of the MPG. One example is Fig. \ref{fig:MPG}(a)
which shows the coreps of the MPG $m'm'm'$: single-valued coreps
in light green background and double-valued coreps in light blue background,
similar to the result of \lstinline!showMLGCorep!. The decomposition
of direct product of MPG coreps can be calculated by \lstinline!MPGCorepDirectProduct[mpg]!
and shown by \lstinline!showMPGCorepDirectProduct[mpg]!. For example,
the direct products for the coreps of MPG 8.5 ($m'm'm')$ are shown
in Fig. \ref{fig:MPG}(c). Furthermore, one can specify the input
coreps by their indexes and even emphasize certain corep(s) in the
results via the \lstinline!"emph"! option. For example, Fig. \ref{fig:MPG}(d)
shows the direct products between two sets of coreps, one set is the
3rd and 5th corep and the other set is all the coreps from the 1st
to the 5th, with emphasizing the 3rd corep (i.e. $B_{2}$) in red
color. 

\section{Applications}

The \textsf{MSGCorep} package has been successfully applied in the
systematic classification of emergent particles, namely the quasiparticles
emerging from various band crossings \citep{Liu_Yao_2022_105_85117__Systematic,Zhang_Yao_2022_105_104426_2112.10479_Encyclopedia,Zhang_Yao_2022___2210.11080_Encyclopedia}.
The properties of a type of emergent particle are determined by the
$\vk\cdot\bm{p}$ model around the band crossing, and the $\vk\cdot\bm{p}$
model can be constructed according to the symmetry constraints imposed
by the small coreps of the degeneracy at the band crossing. In order
to interface with the systematic full data of small coreps provided
by \textsf{MSGCorep} for all k-points and all MSGs, a package called
\textsf{MagneticKP} is developed to construct $\vk\cdot\bm{p}$ models
based on an efficient iterative simplification algorithm \citep{Zhang_Yao_2022___2205.05830_MagneticKP}.
Accordingly, we have systematically investigated the classification
of emergent particles in all the 674 type-III MSGs \citep{Liu_Yao_2022_105_85117__Systematic}
and 517 type-IV MSGs \citep{Zhang_Yao_2022_105_104426_2112.10479_Encyclopedia}
by means of analyzing the $\vk\cdot\bm{p}$ models obtained from \textsf{MagneticKP}
plus \textsf{MSGCorep}. Furthermore, we have also proposed a method
of calculating the small coreps of magnetic subperiodic groups, specifically
all the 528 magnetic layer groups and 394 magnetic rod groups, using
the small coreps data from \textsf{MSGCorep}, and have systematically
classified the possible emergent particles existing in these magnetic
subperiodic groups \citep{Zhang_Yao_2022___2210.11080_Encyclopedia}.

\section{Conclusions}

We have developed a package named \textsf{MSGCorep} to provide convenient
offline access to the complete corep data of all 1651 MSGs in three-dimensional
space. In addition, access to the coreps of 122 MPGs is also supported
by this package. The functionalities of this package include: obtaining
the elements of any MSG, MLG, MPG, and their double groups; calculating
the multiplication, inverse, and power of group elements; obtaining
the small coreps at any k-point and full coreps of any magnetic k-star
for any MSG and showing them in a user-friendly table form; calculating
and showing the decomposition of direct products of full coreps between
any two specified magnetic k-stars; supporting both single-valued
and double-valued coreps; and determining the small coreps of energy
bands. To the best of our knowledge, \textsf{MSGCorep} is the first
package that provides tools to calculate the direct product of full
coreps for any MSG and to determine small coreps of energy bands for
general purpose. In a word, the \textsf{MSGCorep} package is a useful
database and tool set for the MSGs, MPGs, and their coreps, it has
been successfully applied in systematic studies on emergent particles,
and it can play an important role in future studies on the symmetries
in magnetic and nonmagnetic materials.

\section*{Acknowledgments}

The authors thank Wanxiang Feng for helpful discussions. This work
is supported by the National Science Foundation of China (Grant Nos.
12234003, 12274028, 52161135108, 12061131002, and 12004028) and the
National Key R\&D Program of China (Grant No. 2020YFA0308800).

\appendix

\section{\label{sec:matN}The matrix $N$ in Eqs. (\ref{eq:corepa}) and (\ref{eq:corepb})}

Although the BC book provides an equation (7.3.53) to calculate the
matrix $N$, it cannot work in a general case because it depends on
an uncertain matrix $X$. Here we use a definite method to find a
unitary matrix $N$ which relates an unitary irep to its equivalent
unitary irep. Suppose that there are in total $s$ unitary ireps,
namely $\Delta^{p}$ ($p=1,2,\cdots,s$), of the group $H$, and that
$\Delta=\Delta^{q}$. For simplicity, denote $\Delta(A^{-1}mA)^{*}$
as $\bar{\Delta}(m)$ and the task becomes to find a matrix $N$ satisfying
$\bar{\Delta}(m)=N^{-1}\Delta^{q}(m)N$ for $\forall m\in H$, i.e.
\begin{equation}
\bar{\Delta}_{ij}(m)=\sum_{k,l}(N^{-1})_{ik}\Delta_{kl}^{q}(m)N_{lj}=\sum_{k,l}N_{ki}^{*}\Delta_{kl}^{q}(m)N_{lj}.
\end{equation}
Multiply two sides of the above equation by $\Delta_{k'l'}^{p}(m)^{*}$
and then sum over $m$ to get
\begin{align}
\sum_{m\in H}\Delta_{k'l'}^{p}(m)^{*}\bar{\Delta}_{ij}(m) & =\sum_{k,l}N_{ki}^{*}N_{lj}\sum_{m\in H}\Delta_{k'l'}^{p}(m)^{*}\Delta_{kl}^{q}(m)\nonumber \\
 & =\sum_{k,l}N_{ki}^{*}N_{lj}\frac{|H|}{d_{q}}\delta_{pq}\delta_{k'k}\delta_{l'l}=N_{k'i}^{*}N_{l'j}\frac{|H|}{d_{q}}\delta_{pq},
\end{align}
in which the orthogonality theorem of irep matrices is used and $d_{q}$
is the dimension of $\Delta^{q}$. Letting $p=q$, we get
\begin{equation}
N_{ki}^{*}N_{lj}=\frac{d_{q}}{|H|}\sum_{m\in H}\Delta_{kl}^{q}(m)^{*}\bar{\Delta}_{ij}(m).\label{eq:NkiNlj}
\end{equation}
Further letting $l=k$ and $j=i$, we get
\begin{equation}
|N_{ki}|^{2}=\frac{d_{q}}{|H|}\sum_{m\in H}\Delta_{kk}^{q}(m)^{*}\bar{\Delta}_{ii}(m).
\end{equation}
Using the above equation to find a nonzero element of the matrix $N$,
e.g. $N_{1t}=\sqrt{|N_{1t}|^{2}}\ne0$ $(k=1,$ $i=t$), then substitute
$k=1$ and $i=t$ into Eq. (\ref{eq:NkiNlj}) and obtain every matrix
element of $N$ by 
\begin{equation}
N_{lj}=\frac{1}{N_{1t}^{*}}\frac{d_{q}}{|H|}\sum_{m\in H}\Delta_{1l}^{q}(m)^{*}\bar{\Delta}_{tj}(m).
\end{equation}

\section{\label{sec:fixTab}Corrections and adaptions to the tables in the
BC book}

\begin{table}[!tbph]
\caption{\label{tab:fixTab7.4}Corrections and adaptions to the BNS symbols
in BC-Tab. 7.4. The column \textquotedblleft Old\textquotedblright{}
lists the original symbols in the BC book, and the column \textquotedblleft New\textquotedblright{}
lists the symbols we actually use in \textsf{MSGCorep}.}

\centering{}
\begin{tabular}{lllllllllll}
\hline 
\Gape{BNS No.}  & Old  & New &  & BNS No.  & Old  & New  &  & BNS No.  & Old  & New\tabularnewline
\hline 
1.2  & $P1'$  & $\textcolor{green!50!black}{P11'}$  &  & 204.32  & $Im'3$  & $\textcolor{blue!70!cyan}{Im'\bar{3}'}$  &  & 224.110  & $Pn3m$  & $Pn\bar{3}m$\tabularnewline
33.155  & $P_{C}na2_{1}$  & $\textcolor{red}{P_{I}na2_{1}}$  &  & 205.33  & $Pa3$  & $Pa\bar{3}$  &  & 224.111  & $Pn3'm$  & $\textcolor{green!50!black}{Pn\bar{3}m1'}$\tabularnewline
114.276  & $P\bar{4}2_{1}c$  & $\textcolor{red}{P\bar{4}2_{1}c1'}$  &  & 205.34  & $Pa3'$  & $\textcolor{green!50!black}{Pa\bar{3}1'}$  &  & 224.112  & $Pn'3m$  & $\textcolor{blue!70!cyan}{Pn'\bar{3}'m}$\tabularnewline
149.22  & $P31'2$  & $\textcolor{green!50!black}{P3121'}$  &  & 205.35  & $Pa'3$  & $\textcolor{blue!70!cyan}{Pa'\bar{3}'}$  &  & 224.113  & $Pn3m'$  & $Pn\bar{3}m'$\tabularnewline
150.26  & $P321'$  & $\textcolor{green!50!black}{P3211'}$  &  & 205.36  & $P_{I}a3$  & $P_{I}a\bar{3}$  &  & 224.114  & $Pn'3m'$  & $\textcolor{blue!70!cyan}{Pn'\bar{3}'m'}$\tabularnewline
151.30  & $P3_{1}1'2$  & $\textcolor{green!50!black}{P3_{1}121'}$  &  & 206.37  & $Ia3$  & $Ia\bar{3}$  &  & 224.115  & $P_{I}n3m$  & $P_{I}n\bar{3}m$\tabularnewline
152.34  & $P3_{1}21'$  & $\textcolor{green!50!black}{P3_{1}211'}$  &  & 206.38  & $Ia3'$  & $\textcolor{green!50!black}{Ia\bar{3}1'}$  &  & 225.116  & $Fm3m$  & $Fm\bar{3}m$\tabularnewline
153.38  & $P3_{2}1'2$  & $\textcolor{green!50!black}{P3_{2}121'}$  &  & 206.39  & $Ia'3$  & $\textcolor{blue!70!cyan}{Ia'\bar{3}'}$  &  & 225.117  & $Fm3'm$  & $\textcolor{green!50!black}{Fm\bar{3}m1'}$\tabularnewline
154.42  & $P3_{2}21'$  & $\textcolor{green!50!black}{P3_{2}211'}$  &  & 207.41  & $P43'2$  & $\textcolor{green!50!black}{P4321'}$  &  & 225.118  & $Fm'3m$  & $\textcolor{blue!70!cyan}{Fm'\bar{3}'m}$\tabularnewline
156.50  & $P3m1'$  & $\textcolor{green!50!black}{P3m11'}$  &  & 208.45  & $P4_{2}3'2$  & $\textcolor{green!50!black}{P4_{2}321'}$  &  & 225.119  & $Fm3m'$  & $Fm\bar{3}m'$\tabularnewline
157.54  & $P31'm$  & $\textcolor{green!50!black}{P31m1'}$  &  & 209.49  & $F43'2$  & $\textcolor{green!50!black}{F4321'}$  &  & 225.120  & $Fm'3m'$  & $\textcolor{blue!70!cyan}{Fm'\bar{3}'m'}$\tabularnewline
158.58  & $P3c1'$  & $\textcolor{green!50!black}{P3c11'}$  &  & 210.53  & $F4_{1}3'2$  & $\textcolor{green!50!black}{F4_{1}321'}$  &  & 225.121  & $F_{s}m3m$  & $F_{s}m\bar{3}m$\tabularnewline
159.62  & $P31'c$  & $\textcolor{green!50!black}{P31c1'}$  &  & 211.57  & $I43'2$  & $\textcolor{green!50!black}{I4321'}$  &  & 226.122  & $Fm3c$  & $Fm\bar{3}c$\tabularnewline
162.74  & $P\bar{3}1'm$  & $\textcolor{green!50!black}{P\bar{3}1m1'}$  &  & 212.60  & $P4_{3}3'2$  & $\textcolor{green!50!black}{P4_{3}321'}$  &  & 226.123  & $Fm3'c$  & $\textcolor{green!50!black}{Fm\bar{3}c1'}$\tabularnewline
163.80  & $P\bar{3}1'c$  & $\textcolor{green!50!black}{P\bar{3}1c1'}$  &  & 213.64  & $P4_{1}3'2$  & $\textcolor{green!50!black}{P4_{1}321'}$  &  & 226.124  & $Fm'3c$  & $\textcolor{blue!70!cyan}{Fm'\bar{3}'c}$\tabularnewline
164.86  & $P\bar{3}m1'$  & $\textcolor{green!50!black}{P\bar{3}m11'}$  &  & 214.68  & $I4_{1}3'2$  & $\textcolor{green!50!black}{I4_{1}321'}$  &  & 226.125  & $Fm3c'$  & $Fm\bar{3}c'$\tabularnewline
165.92  & $P\bar{3}c1'$  & $\textcolor{green!50!black}{P\bar{3}c11'}$  &  & 215.71  & $P\bar{4}3'm$  & $\textcolor{green!50!black}{P\bar{4}3m1'}$  &  & 226.126  & $Fm'3c'$  & $\textcolor{blue!70!cyan}{Fm'\bar{3}'c'}$\tabularnewline
195.2  & $P23'$  & $\textcolor{green!50!black}{P231'}$  &  & 216.75  & $F\bar{4}3'm$  & $\textcolor{green!50!black}{F\bar{4}3m1'}$  &  & 226.127  & $F_{s}m3c$  & $F_{s}m\bar{3}c$\tabularnewline
196.5  & $F23'$  & $\textcolor{green!50!black}{F231'}$  &  & 217.79  & $I\bar{4}3'm$  & $\textcolor{green!50!black}{I\bar{4}3m1'}$  &  & 227.128  & $Fd3m$  & $Fd\bar{3}m$\tabularnewline
197.8  & $I23'$  & $\textcolor{green!50!black}{I231'}$  &  & 218.82  & $P\bar{4}3'n$  & $\textcolor{green!50!black}{P\bar{4}3n1'}$  &  & 227.129  & $Fd3'm$  & $\textcolor{green!50!black}{Fd\bar{3}m1'}$\tabularnewline
198.10  & $P2_{1}3'$  & $\textcolor{green!50!black}{P2_{1}31'}$  &  & 219.86  & $F\bar{4}3'c$  & $\textcolor{green!50!black}{F\bar{4}3c1'}$  &  & 227.130  & $Fd'3m$  & $\textcolor{blue!70!cyan}{Fd'\bar{3}'m}$\tabularnewline
199.13  & $I2_{1}3'$  & $\textcolor{green!50!black}{I2_{1}31'}$  &  & 220.90  & $I\bar{4}3'd$  & $\textcolor{green!50!black}{I\bar{4}3d1'}$  &  & 227.131  & $Fd3m'$  & $Fd\bar{3}m'$\tabularnewline
200.14  & $Pm3$  & $Pm\bar{3}$  &  & 221.92  & $Pm3m$  & $Pm\bar{3}m$  &  & 227.132  & $Fd'3m'$  & $\textcolor{blue!70!cyan}{Fd'\bar{3}'m'}$\tabularnewline
200.15  & $Pm3'$  & $\textcolor{green!50!black}{Pm\bar{3}1'}$  &  & 221.93  & $Pm3'm$  & $\textcolor{green!50!black}{Pm\bar{3}m1'}$  &  & 227.133  & $F_{s}d3m$  & $F_{s}d\bar{3}m$\tabularnewline
200.16  & $Pm'3$  & $\textcolor{blue!70!cyan}{Pm'\bar{3}'}$  &  & 221.94  & $Pm'3m$  & $\textcolor{blue!70!cyan}{Pm'\bar{3}'m}$  &  & 228.134  & $Fd3c$  & $Fd\bar{3}c$\tabularnewline
200.17  & $P_{I}m3$  & $P_{I}m\bar{3}$  &  & 221.95  & $Pm3m'$  & $Pm\bar{3}m'$  &  & 228.135  & $Fd3'c$  & $\textcolor{green!50!black}{Fd\bar{3}c1'}$\tabularnewline
201.18  & $Pn3$  & $Pn\bar{3}$  &  & 221.96  & $Pm'3m'$  & $\textcolor{blue!70!cyan}{Pm'\bar{3}'m'}$  &  & 228.136  & $Fd'3c$  & $\textcolor{blue!70!cyan}{Fd'\bar{3}'c}$\tabularnewline
201.19  & $Pn3'$  & $\textcolor{green!50!black}{Pn\bar{3}1'}$  &  & 221.97  & $P_{I}m3m$  & $P_{I}m\bar{3}m$  &  & 228.137  & $Fd3c'$  & $Fd\bar{3}c'$\tabularnewline
201.20  & $Pn'3$  & $\textcolor{blue!70!cyan}{Pn'\bar{3}'}$  &  & 222.98  & $Pn3n$  & $Pn\bar{3}n$  &  & 228.138  & $Fd'3c'$  & $\textcolor{blue!70!cyan}{Fd'\bar{3}'c'}$\tabularnewline
201.21  & $P_{I}n3$  & $P_{I}n\bar{3}$  &  & 222.99  & $Pn3'n$  & $\textcolor{green!50!black}{Pn\bar{3}n1'}$  &  & 228.139  & $F_{s}d3c$  & $F_{s}d\bar{3}c$\tabularnewline
202.22  & $Fm3$  & $Fm\bar{3}$  &  & 222.100  & $Pn'3n$  & $\textcolor{blue!70!cyan}{Pn'\bar{3}'n}$  &  & 229.140  & $Im3m$  & $Im\bar{3}m$\tabularnewline
202.23  & $Fm3'$  & $\textcolor{green!50!black}{Fm\bar{3}1'}$  &  & 222.101  & $Pn3n'$  & $Pn\bar{3}n'$  &  & 229.141  & $Im3'm$  & $\textcolor{green!50!black}{Im\bar{3}m1'}$\tabularnewline
202.24  & $Fm'3$  & $\textcolor{blue!70!cyan}{Fm'\bar{3}'}$  &  & 222.102  & $Pn'3n'$  & $\textcolor{blue!70!cyan}{Pn'\bar{3}'n'}$  &  & 229.142  & $Im'3m$  & $\textcolor{blue!70!cyan}{Im'\bar{3}'m}$\tabularnewline
202.25  & $F_{s}m3$  & $F_{s}m\bar{3}$  &  & 222.103  & $P_{I}n3n$  & $P_{I}n\bar{3}n$  &  & 229.143  & $Im3m'$  & $Im\bar{3}m'$\tabularnewline
203.26  & $Fd3$  & $Fd\bar{3}$  &  & 223.104  & $Pm3n$  & $Pm\bar{3}n$  &  & 229.144  & $Im'3m'$  & $\textcolor{blue!70!cyan}{Im'\bar{3}'m'}$\tabularnewline
203.27  & $Fd3'$  & $\textcolor{green!50!black}{Fd\bar{3}1'}$  &  & 223.105  & $Pm3'n$  & $\textcolor{green!50!black}{Pm\bar{3}n1'}$  &  & 230.145  & $Ia3d$  & $Ia\bar{3}d$\tabularnewline
203.28  & $Fd'3$  & $\textcolor{blue!70!cyan}{Fd'\bar{3}'}$  &  & 223.106  & $Pm'3n$  & $\textcolor{blue!70!cyan}{Pm'\bar{3}'n}$  &  & 230.146  & $Ia3'd$  & $\textcolor{green!50!black}{Ia\bar{3}d1'}$\tabularnewline
203.29  & $F_{s}d3$  & $F_{s}d\bar{3}$  &  & 223.107  & $Pm3n'$  & $Pm\bar{3}n'$  &  & 230.147  & $Ia'3d$  & $\textcolor{blue!70!cyan}{Ia'\bar{3}'d}$\tabularnewline
204.30  & $Im3$  & $Im\bar{3}$  &  & 223.108  & $Pm'3n'$  & $\textcolor{blue!70!cyan}{Pm'\bar{3}'n'}$  &  & 230.148  & $Ia3d'$  & $Ia\bar{3}d'$\tabularnewline
204.31  & $Im3'$  & $\textcolor{green!50!black}{Im\bar{3}1'}$  &  & 223.109  & $P_{I}m3n$  & $P_{I}m\bar{3}n$  &  & 230.149  & $Ia'3d'$  & $\textcolor{blue!70!cyan}{Ia'\bar{3}'d'}$\tabularnewline
\hline 
\end{tabular}
\end{table}

\textbf{BC-Tab. 7.4}: As stated in the BC book, BC-Tab. 7.4 is basically
the same as the original table given by Belov, Neronova, and Smirnova
in 1957 \citep{Belov_Smirnova_1957_2_311__Shubnikov}. To be consistent
with the BNS symbols given in the latest monograph about MSGs by Litvin
\citep{Litvin2013MGT}, we have made some adaptions to BC-Tab. 7.4.
Firstly, we make all the symbols of type-II MSGs be the corresponding
ones of type-I MSGs followed by $1'$. Not all the original type-II
symbols in BC-Tab. 7.4 obey this rule, such as some trigonal MSGs
(e.g. 149.22 $P31'2$) and all cubic MSGs (e.g. 207.41 $P43'2$).
New symbols of this type are shown in green color in Table \ref{tab:fixTab7.4}.
Secondly, all the MSG symbols derived from the PG symbols $m3$ ($T_{h}$)
and $m3m$ ($O_{h}$) should be updated, because $m3$ and $m3m$
have been changed to $m\bar{3}$ and $m\bar{3}m$ respectively for
tens of years. This change from 3 to $\bar{3}$ involves all the MSGs
with numbers $n.m$, $n\in[200,206]\cup[221,230]$. Especially, there
are 27 type-III MSGs for which $\bar{3}$ should be further primed
after the change, which are shown in blue color in Table \ref{tab:fixTab7.4}.
Finally, there are two typos and the corrections are shown in red
color in Table \ref{tab:fixTab7.4}.

\textbf{BC-Tab. 7.2}: See Table \ref{tab:fixTab7.2} for the corrections
to BC-Tab. 7.2.

\begin{table}[!tbph]
\caption{\label{tab:fixTab7.2}Corrections to the \textquotedblleft colored
generating elements (CGE)\textquotedblright{} in BC-Tab. 7.2. The column
\textquotedblleft SymStd\textquotedblright{} lists the standard BNS
symbols. Here we also list the BNS symbols conforming to the BC orientation
for comparison, if they are different from the standard symbols.}

\centering{}
\begin{tabular}{lllllllll}
\hline 
BNS No.  & SymStd  & SymBC  & \Gape{CGE}  & $\ $ & BNS No.  & SymStd  & SymBC  & CGE \tabularnewline
\hline 
28.89  & $Pm'a2'$  & $Pbm'2'$  & $\sigma_{y}$  &  & 57.380  & $Pbc'm$  & $Pb'ma$  & ${C}_{2y},{I}$\tabularnewline
28.90  & $Pma'2'$  & $Pb'm2'$  & $\sigma_{x}$  &  & 57.381  & $Pbcm'$  & $Pbm'a$  & ${C}_{2x},{I}$\tabularnewline
29.101  & $Pc'a2_{1}'$  & $Pbc'2_{1}'$  & $\sigma_{y}$  &  & 57.382  & $Pb'c'm$  & $Pb'ma'$  & ${C}_{2x}$\tabularnewline
29.102  & $Pca'2_{1}'$  & $Pb'c2_{1}'$  & $\sigma_{x}$  &  & 57.383  & $Pbc'm'$  & $Pb'm'a$  & ${C}_{2x},{C}_{2y}$\tabularnewline
31.125  & $Pm'n2_{1}'$  & $Pnm'2_{1}'$  & $\sigma_{y}$  &  & 57.384  & $Pb'cm'$  & $Pbm'a'$  & ${C}_{2y}$\tabularnewline
31.126  & $Pmn'2_{1}'$  & $Pn'm2_{1}'$  & $\sigma_{x}$  &  & 60.420  & $Pbc'n$  & $Pcnb'$  & ${C}_{2x},{C}_{2y},{I}$\tabularnewline
33.146  & $Pn'a2_{1}'$  & $Pbn'2_{1}'$  & $\sigma_{y}$  &  & 60.421  & $Pbcn'$  & $Pcn'b$  & ${C}_{2x},{I}$\tabularnewline
33.147  & $Pna'2_{1}'$  & $Pb'n2_{1}'$  & $\sigma_{x}$  &  & 60.422  & $Pb'c'n$  & $Pc'nb'$  & ${C}_{2x}$\tabularnewline
36.174  & $Cm'c2_{1}'$  & $Ccm'2_{1}'$  & $\sigma_{y}$  &  & 60.424  & $Pb'cn'$  & $Pc'n'b$  & ${C}_{2x},{C}_{2y}$\tabularnewline
36.175  & $Cmc'2_{1}'$  & $Cc'm2_{1}'$  & $\sigma_{x}$  &  & 160.67  & $R3m'$  &  & $\sigma_{d1}$\tabularnewline
38.189  & $Am'm2'$  & $Cm2'm'$  & $\sigma_{z}$  &  & 161.71  & $R3c'$  &  & $\sigma_{d1}$\tabularnewline
38.190  & $Amm'2'$  & $Cm'2'm$  & $\sigma_{x}$  &  & 166.100  & $R\bar{3}'m'$  &  & ${S}_{6}^{+},\sigma_{d1}$\tabularnewline
38.191  & $Am'm'2$  & $Cm'2m'$  & $\sigma_{z},\sigma_{x}$  &  & 166.101  & $R\bar{3}m'$  &  & $\sigma_{d1}$\tabularnewline
39.197  & $Ab'm2'$  & $Cm2'a'$  & $\sigma_{z}$  &  & 167.106  & $R\bar{3}'c'$  &  & ${S}_{6}^{+},\sigma_{d1}$\tabularnewline
39.198  & $Abm'2'$  & $Cm'2'a$  & $\sigma_{x}$  &  & 167.107  & $R\bar{3}c'$  &  & $\sigma_{d1}$\tabularnewline
39.199  & $Ab'm'2$  & $Cm'2a'$  & $\sigma_{z},\sigma_{x}$  &  & 177.151  & $P6'2'2$  &  & ${C}_{6}^{+}$\tabularnewline
40.205  & $Am'a2'$  & $Cc2'm'$  & $\sigma_{z}$  &  & 177.152  & $P6'22'$  &  & ${C}_{6}^{+},{C}_{21}'$\tabularnewline
40.206  & $Ama'2'$  & $Cc'2'm$  & $\sigma_{x}$  &  & 178.157  & $P6_{1}'2'2$  &  & ${C}_{6}^{+}$\tabularnewline
40.207  & $Am'a'2$  & $Cc'2m'$  & $\sigma_{z},\sigma_{x}$  &  & 178.158  & $P6_{1}'22'$  &  & ${C}_{6}^{+},{C}_{21}'$\tabularnewline
41.213  & $Ab'a2'$  & $Cc2'a'$  & $\sigma_{z}$  &  & 179.163  & $P6_{5}'2'2$  &  & ${C}_{6}^{+}$\tabularnewline
41.214  & $Aba'2'$  & $Cc'2'a$  & $\sigma_{x}$  &  & 179.164  & $P6_{5}'22'$  &  & ${C}_{6}^{+},{C}_{21}'$\tabularnewline
41.215  & $Ab'a'2$  & $Cc'2a'$  & $\sigma_{z},\sigma_{x}$  &  & 180.169  & $P6_{2}'2'2$  &  & ${C}_{6}^{+}$\tabularnewline
46.243  & $Im'a2'$  & $Ibm'2'$  & $\sigma_{y}$  &  & 180.170  & $P6_{2}'22'$  &  & ${C}_{6}^{+},{C}_{21}'$\tabularnewline
46.244  & $Ima'2'$  & $Ib'm2'$  & $\sigma_{x}$  &  & 181.175  & $P6_{4}'2'2$  &  & ${C}_{6}^{+}$\tabularnewline
51.291  & $Pm'ma$  & $Pcmm'$  & ${C}_{2x},{C}_{2y},{I}$  &  & 181.176  & $P6_{4}'22'$  &  & ${C}_{6}^{+},{C}_{21}'$\tabularnewline
51.293  & $Pmma'$  & $Pc'mm$  & ${C}_{2y},{I}$  &  & 182.181  & $P6_{3}'2'2$  &  & ${C}_{6}^{+}$\tabularnewline
51.294  & $Pm'm'a$  & $Pcm'm'$  & ${C}_{2y}$  &  & 182.182  & $P6_{3}'22'$  &  & ${C}_{6}^{+},{C}_{21}'$\tabularnewline
51.295  & $Pmm'a'$  & $Pc'm'm$  & ${C}_{2x},{C}_{2y}$  &  & 183.187  & $P6'm'm$  &  & ${C}_{6}^{+},\sigma_{v1}$\tabularnewline
52.308  & $Pnn'a$  & $Pnan'$  & ${C}_{2x},{C}_{2y},{I}$  &  & 183.188  & $P6'mm'$  &  & ${C}_{6}^{+}$\tabularnewline
52.309  & $Pnna'$  & $Pna'n$  & ${C}_{2x},{I}$  &  & 184.193  & $P6'c'c$  &  & ${C}_{6}^{+},\sigma_{v1}$\tabularnewline
52.310  & $Pn'n'a$  & $Pn'an'$  & ${C}_{2x}$  &  & 184.194  & $P6'cc'$  &  & ${C}_{6}^{+}$\tabularnewline
52.312  & $Pn'na'$  & $Pn'a'n$  & ${C}_{2x},{C}_{2y}$  &  & 185.199  & $P6_{3}'c'm$  &  & ${C}_{6}^{+},\sigma_{v1}$\tabularnewline
53.323  & $Pm'na$  & $Pnm'b$  & ${C}_{2x},{I}$  &  & 185.200  & $P6_{3}'cm'$  &  & ${C}_{6}^{+}$\tabularnewline
53.324  & $Pmn'a$  & $Pn'mb$  & ${C}_{2y},{I}$  &  & 186.205  & $P6_{3}'m'c$  &  & ${C}_{6}^{+},\sigma_{v1}$\tabularnewline
53.327  & $Pmn'a'$  & $Pn'mb'$  & ${C}_{2x}$  &  & 186.206  & $P6_{3}'mc'$  &  & ${C}_{6}^{+}$\tabularnewline
53.328  & $Pm'na'$  & $Pnm'b'$  & ${C}_{2y}$  &  & 191.236  & $P6'/mm'm$  &  & ${C}_{6}^{+},{C}_{21}',{I}$\tabularnewline
54.339  & $Pc'ca$  & $Pcaa'$  & ${C}_{2x},{C}_{2y},{I}$  &  & 191.237  & $P6'/mmm'$  &  & ${C}_{6}^{+},{I}$\tabularnewline
54.341  & $Pcca'$  & $Pc'aa$  & ${C}_{2y},{I}$  &  & 191.238  & $P6'/m'm'm$  &  & ${C}_{6}^{+}$\tabularnewline
54.342  & $Pc'c'a$  & $Pca'a'$  & ${C}_{2y}$  &  & 191.239  & $P6'/m'mm'$  &  & ${C}_{6}^{+},{C}_{21}'$\tabularnewline
54.343  & $Pcc'a'$  & $Pc'a'a$  & ${C}_{2x},{C}_{2y}$  &  & 192.246  & $P6'/mc'c$  &  & ${C}_{6}^{+},{C}_{21}',{I}$\tabularnewline
57.379  & $Pb'cm$  & $Pbma'$  & ${C}_{2x},{C}_{2y},{I}$  &  & 192.247  & $P6'/mcc'$  &  & ${C}_{6}^{+},{I}$\tabularnewline
\hline 
\end{tabular}
\end{table}

\textbf{BC-Tab. 7.3}: See Table \ref{tab:BWlatt} for the supplement
to BC-Tab. 7.3.

\begin{table}[!th]
\caption{\label{tab:BWlatt}Black-white lattices which should be added to BC-Tab.
7.3 to make it complete.}

{\renewcommand{\arraystretch}{1.3}
\begin{centering}
\begin{tabular}{lccl}
\hline 
Crystal system & Ordinary lattice & Black-white lattice & $\bm{t}_{0}$\tabularnewline
\hline 
Monoclinic & $P(\Gamma_{m})$ & $P_{c}$ & $\frac{1}{2}\bm{t}_{1}$\tabularnewline
 &  & $P_{A}$ & $\frac{1}{2}(\bm{t}_{1}+\bm{t}_{3})$\tabularnewline
Orthorhombic & $P(\Gamma_{o})$ & $P_{b}$ & $\frac{1}{2}\bm{t}_{1}$\tabularnewline
 &  & $P_{B}$ & $\frac{1}{2}(\bm{t}_{2}+\bm{t}_{3})$\tabularnewline
Orthorhombic & $I(\Gamma_{o}^{v})$ & $I_{a}$ & $\frac{1}{2}(\bm{t}_{1}+\bm{t}_{3})$\tabularnewline
 &  & $I_{b}$ & $\frac{1}{2}(\bm{t}_{2}+\bm{t}_{3})$\tabularnewline
\hline 
\end{tabular}
\par\end{centering}
}

\end{table}

\section{\label{sec:diffSym}Differences of MSG symbols between Litvin and
ISO-MAG}

The standard MSG symbols of both BNS and OG types here we use are
consistent with those used by Litvin \citep{Litvin2013MGT}. But they
are different from those used by ISO-MAG \citep{iso-mag} and BCS
\citep{BCSweb} for some MSGs. We list these differences in Tables
\ref{tab:diffsymOG} and \ref{tab:diffsymBNS} for reference so as
not to result in misunderstanding in certain cases. The version 1.3
(Jan 2022) of ISO-MAG is consistent with BCS, and both Tables \ref{tab:diffsymOG}
and \ref{tab:diffsymBNS} show their differences from Litvin. After
our communication to one author of ISO-MAG about the differences,
the version 1.4 (May 2022) of ISO-MAG changed the seven OG symbols
in Table \ref{tab:diffsymOG} to those of Litvin. However, the twenty-one
BNS symbols of ISO-MAG in Table \ref{tab:diffsymBNS} keep the same
from version 1.3 to 1.4 because the authors want to keep them consistent
with BCS. 

\begin{table}[!th]
\caption{\label{tab:diffsymOG}Differences of OG symbols between Litvin and
ISO-MAG of version 1.3 as well as BCS. These symbols of ISO-MAG have
been changed to be the same with those of Litvin in version 1.4.}

\begin{centering}
\begin{tabular}{lll}
\hline 
\Gape{OG No.} & Litvin & ISO-MAG (v1.3)\tabularnewline
\hline 
16.6.104 & $P_{F}222$ & $P_{I}222$\tabularnewline
25.9.163 & $P_{F}mm2$ & $P_{I}mm2$\tabularnewline
34.5.235 & $P_{F}nn2$ & $P_{I}nn2$\tabularnewline
47.8.354 & $P_{F}mmm$ & $P_{I}mmm$\tabularnewline
48.6.363 & $P_{F}nnn$ & $P_{I}nnn$\tabularnewline
153.4.1270 & $P_{2c}3_{1}12'$ & $P_{2c}3_{1}12$\tabularnewline
154.4.1274 & $P_{2c}3_{1}2'1$ & $P_{2c}3_{1}21$\tabularnewline
\hline 
\end{tabular}
\par\end{centering}
\end{table}

\begin{table}[!th]
\caption{\label{tab:diffsymBNS}Differences of BNS symbols between Litvin and
ISO-MAG as well as BCS. These symbols of ISO-MAG did not change from
version 1.3 to 1.4.}

\centering{}%
\begin{tabular}{lllllllllll}
\hline 
\Gape{BNS No.} & Litvin & ISO-MAG &  & BNS No. & Litvin & ISO-MAG &  & BNS No. & Litvin & ISO-MAG\tabularnewline
\hline 
17.13 & $P_{A}222_{1}$ & $P_{B}222_{1}$ &  & 39.201 & $A_{c}bm2$ & $A_{b}bm2$ &  & 49.274 & $P_{A}ccm$ & $P_{B}ccm$\tabularnewline
18.20 & $P_{a}2_{1}2_{1}2$ & $P_{b}2_{1}2_{1}2$ &  & 39.202 & $A_{C}bm2$ & $A_{B}bm2$ &  & 56.372 & $P_{a}ccn$ & $P_{b}ccn$\tabularnewline
18.22 & $P_{A}2_{1}2_{1}2$ & $P_{B}2_{1}2_{1}2$ &  & 40.209 & $A_{c}ma2$ & $A_{b}ma2$ &  & 58.402 & $P_{A}nnm$ & $P_{B}nnm$\tabularnewline
19.28 & $P_{a}2_{1}2_{1}2_{1}$ & $P_{c}2_{1}2_{1}2_{1}$ &  & 40.210 & $A_{C}ma2$ & $A_{B}ma2$ &  & 59.412 & $P_{a}mmn$ & $P_{b}mmn$\tabularnewline
32.140 & $P_{a}ba2$ & $P_{b}ba2$ &  & 41.217 & $A_{c}ba2$ & $A_{b}ba2$ &  & 59.414 & $P_{A}mmn$ & $P_{B}mmn$\tabularnewline
38.193 & $A_{c}mm2$ & $A_{b}mm2$ &  & 41.218 & $A_{C}ba2$ & $A_{B}ba2$ &  & 72.547 & $I_{a}bam$ & $I_{b}bam$\tabularnewline
38.194 & $A_{C}mm2$ & $A_{B}mm2$ &  & 48.262 & $P_{a}nnn$ & $P_{c}nnn$ &  & 74.562 & $I_{a}mma$ & $I_{b}mma$\tabularnewline
\hline 
\end{tabular}
\end{table}


\begin{thebibliography}{61}
\bibitem[{typ(????)}]{typeI}
\bibinfo{title}{{When only geometry structures of crystals are considered,
  type-I MSGs can describe them. But when magnetic structures are considered, a
  nonmagnetic material has zero magnetic moments everywhere and hence $\tr$
  symmetry exists. The MSGs containing $\tr$ are of type-II not type-I.}}
\end{thebibliography}

\end{document}